\documentclass[titlepage,12pt,subeqn]{utarticle}
\usepackage{amsmath,amssymb,amscd}
\usepackage{color}
\usepackage{latexsym}
\usepackage{ifthen}
\usepackage[xdvi]{graphics}
\usepackage[]{hyperref} 
\usepackage{bbm}

\numberwithin{equation}{section}

\newcommand{\ic}{{\rm i}}

\newcommand{\be}{\begin{equation}}
\newcommand{\ee}{\end{equation}}
\newcommand{\bea}{\begin{eqnarray}}
\newcommand{\eea}{\end{eqnarray}}
\newcommand{\al}{\alpha}
\renewcommand{\d}{\delta}

\newcommand{\G}{\Gamma}
\newcommand{\g}{\gamma}
\renewcommand{\k}{\kappa}

\newcommand{\m}{\mu}

\newcommand{\Om}{\Omega}
\newcommand{\om}{\omega}
\newcommand{\s}{\sigma}

\newcommand{\hlf}{\frac{1}{2}}

\newcommand{\non}{\nonumber}
\newcommand{\N}{\mathcal{N}}
\newcommand{\p}{\partial}

\newcommand{\R}{\mathbb{R}}
\newcommand{\rr}{\rightarrow}
\newcommand{\w}{\wedge}
\newcommand{\Z}{\mathbb{Z}}

\renewcommand{\Im}{\operatorname{Im}}

\renewcommand{\Re}{\operatorname{Re}}

\newcommand{\SO}{\operatorname{SO}}
\newcommand{\Sp}{\operatorname{Sp}}

\newcommand{\U}{\operatorname{U}}
\newcommand{\lp}{\left(}
\newcommand{\rp}{\right)}
\newcommand{\ls}{\left[}
\newcommand{\rs}{\right]}

\long\def\del#1\enddel{}        \long\def\new#1\endnew{{\bf #1}}

\def\ifundefined#1{\expandafter\ifx\csname#1\endcsname\relax}
\ifundefined{draftmode} {} \else        
   \newcount\HOUR  \HOUR=\the\time \divide\HOUR by 60    \multiply \HOUR by 60
   \newcount\MIN   \MIN=\the\time  \advance\MIN by -\HOUR   \divide\HOUR by 60
   \ifnum   \MIN>9  \def\printTIME{{\it\the\HOUR\,:\,\the\MIN}}
         \else   \def\printTIME{{\it\the\HOUR\,:\,0\the\MIN}} \fi 
   \renewcommand{\thepage}{\hbox to9pt{\hss{draft version~ \it \draftmode
        } ~ ~ --- ~ ~ } {\bf\arabic{page}}
        \hbox to9pt{ ~ ~ --- ~ ~ {generated \it\today} ~ ~at~ \printTIME\hss}}
\fi

\begin{document}
\preprint{
UTTG-07-07 \\
}
\title{D-Terms from Generalized NS-NS Fluxes in Type II}

\author{ {\sc Daniel Robbins} and {\sc Timm Wrase}
     \oneaddress{
      Theory Group, Department of Physics,\\
      University of Texas at Austin,\\
      Austin, TX 78712, USA \\
      {~}\\
      \email{robbins@zippy.ph.utexas.edu}\\
      \email{wrase@zippy.ph.utexas.edu}\\
      }
}

\date{September 13, 2007}

\Abstract{Orientifolds of type II string theory admit a certain set of generalized NS-NS fluxes, including not only the three-form field strength $H$, but also metric and non-geometric fluxes, which are related to $H$ by T-duality.  We describe in general how these fluxes appear as parameters of an effective $\mathcal{N}=1$ supergravity theory in four dimensions, and in particular how certain generalized NS-NS fluxes can act as charges for R-R axions, leading to D-term contributions to the effective scalar potential.  We illustrate these phenomena in type IIB with the example of a certain orientifold of $T^6/\Z_4$.}


\maketitle
\section{Introduction}

Contact with four-dimensional physics is the goal of much recent research exploring the space of string theory vacua.  Of particular interest are those vacua which can be described within the formalism of $\mathcal{N}=1$ supergravity in four dimensions.  Once we have an effective theory in this language, we can explore our field space, looking for extrema of the scalar potential that are either supersymmetric or in which supersymmetry is spontaneously broken.  We are especially interested in solutions that do not have any flat directions, i.e. all moduli are stabilized, and in regions of field space in which inflation can occur, perhaps ideally a sort of hybrid inflation scenario in which one eventually exits the inflationary region and rolls down to a metastable de Sitter minimum.  We also, of course, want to be able to understand any relevant corrections to the effective description in each of these cases.

This impressive list of demands we have of our effective theory will likely require us to find examples in which the scalar potential has as rich a structure as possible.  In the present work we will be focusing on Calabi-Yau orientifolds of type II string theory.  Starting just with such a compactification, one finds the correct supersymmetry in four-dimensions, but there is no superpotential generated and no scalar fields are charged.  Additionally there are typically R-R tadpoles generated by the orientifold planes.  To alleviate these problems one can add D-branes and fluxes to these constructions.  In particular, fluxes, which are expectation values for certain R-R and NS-NS field strengths which arise in the ten-dimensional theory, can generate a superpotential in the effective theory, as well as contributions to the tadpole constraints which can in principle cancel the orientifold contributions (D-branes can also be used for the latter purpose).  In one case, that of an O5/O9 type orientifold of IIB string theory with a flux of the NS-NS three form $H$, one scalar field (from the period of $C_6$ over the internal space) can get charged under the $\U(1)$ gauge group obtained by reducing $C_4$ against the dual three-form to $H$~\cite{Grimm:2004uq}, and this can lead to a simple D-term contribution to the scalar potential in this case.

However, it turns out that even the class of orientifolds with the usual fluxes in type II is still not as rich as we would like.  For instance, in type IIA one can use the flux induced superpotential~\cite{Gukov:1999ya} to stabilize all moduli in some special cases~\cite{DeWolfe:2005uu}, but in general some axions will remain unfixed~\cite{DeWolfe:2005uu,Ihl:2006pp}.  In type IIB the situation is even worse, as the perturbative scalar potential is generically independent of the overall volume modulus.  Nonperturbative effects can sometimes be used to fix all moduli, but in general it is very difficult to lift all of the flat directions.  In all the cases it seems very difficult to find metastable de Sitter minima of the potential and to find regions where slow-roll inflation can occur~\cite{Hertzberg:2007ke}.

For these reasons, it is important to consider what other ingredients might be added to these string theory compactifications which might enrich the structure of the effective four dimensional theory.  T-duality provides a hint.  By performing a T-duality along a circle that has a non-trivial $H$-flux component (i.e. if the circle isometry contracted with $H$ is non-zero) one generates a new solution in which some components of $H$ have been exchanged for non-constant metric components.  These twists of the internal space metric can be represented by components $\om^i_{jk}$ (analogous to $H_{ijk}$) and are usually called metric (or sometimes geometric) fluxes.  By reducing the ten-dimensional supergravity action along this new space, one can learn how these new objects enter into the effective theory.  Like $H$-flux and R-R fluxes, they enter as parameters  in the superpotential and tadpole constraints, and can also charge some of the scalar fields in the theory leading to D-terms~\cite{Ihl:2007ah}.  One can sometimes perform further T-dualities and obtain a space that is no longer globally a manifold, but rather should be thought of as a string compactification whose transition functions lie in a stringy duality group, e.g. $\SO(6,6;\Z)$ for a six-torus.  The twists in this case are described by so-called nongeometric fluxes $Q^{ij}_k$.  At the level of effective theory one can also include another type of non-geometric flux, $R^{ijk}$, which would be purportedly dual to $H_{ijk}$ by T-dualizing all three legs, but these (and a subset of the more conventional fluxes) seem difficult to construct from a ten-dimensional perspective.  However, since we will be mainly concerned with the effective field theories in the present work, we can easily include the full set of plausible general NS-NS fluxes.  The way that all of these fluxes appear in the four-dimensional effective theory can be deduced by T-duality arguments~\cite{Shelton:2005cf,Shelton:2006fd,Aldazabal:2006up}.  For a recent review of these nongeometric fluxes, see~\cite{Wecht:2007wu}, and references therein.  For a more careful exposition of the approach we will be following here, please refer to~\cite{Ihl:2007ah}.

It turns out that adding these extra ingredients really does alleviate some of the problems mentioned above~\cite{Camara:2005dc,Shelton:2005cf,Shelton:2006fd,Aldazabal:2006up,Ihl:2007ah}.  Perturbative moduli stabilization is improved, so that for instance in type IIA all of the moduli, including all axions, can be stabilized, while in type IIB we can generate potentials for all moduli, including the volume modulus (see also the related nongeometric construction~\cite{Becker:2006ks}).  Additionally, in IIA it was shown~\cite{Ihl:2007ah} (see also~\cite{Koerber:2007xk}) that D-terms are generated by some of the metric and non-geometric fluxes.  In section~\ref{IIBCase} we will see that we can also generate D-terms, both in O3/O7 models as well as O5/O9 models.  The D-term found in~\cite{Grimm:2004uq} is a special case of the latter.  Since the D-term contribution to the scalar potential is always non-negative, one might hope that their presence will improve the prospects of finding stabilized de Sitter minima and candidate regions for slow-roll inflation.

The goal of this paper is to present a general formalism for the effective $\mathcal{N}=1$ four-dimensional supergravity with all these generalized fluxes included, and in particular to explain how some of the fluxes lead to charged scalars and hence D-terms.

This idea of fluxes (either ordinary fluxes or the generalized fluxes we are considering here) playing the role of charging certain scalar fields is actually not very unusual in the subject of string compactifications\footnote{We would like to thank Simeon Hellerman for pointing this out to us.}.  Besides the examples discussed here, we will mention one other example.  In type IIA string theory on a Calabi-Yau three-fold with no orientifold (so that we have $\mathcal{N}=2$ in four dimensions), we have a universal NS-NS axion $a$, obtained by dualizing the four-dimensional part of the NS-NS $B$-field (note that an orientifold projects out this axion).  In~\cite{Polchinski:1995sm}, and later refined in~\cite{Louis:2002ny}, it is explained that turning on background R-R fluxes on the internal space lead to electric and magnetic charges for $a$.  In some sense the structures we will be describing are very analogous, but with the roles of NS-NS and R-R fields reversed, i.e. it will be the (generalized) NS-NS fluxes which act as charges for R-R axions (see also~\cite{D'Auria:2007ay,Ihl:2007ah,Palti:2007pm}.

The plan of the paper is as follows.  In section~\ref{IIACase} we will review the story in the case of IIA orientifolds as previously derived in~\cite{Ihl:2007ah}.  This includes a review of the $\mathcal{N}=1$ SUGRA formalism and a description of our generalized NS-NS fluxes, both of which largely carry over to the IIB case.  Section~\ref{IIBCase} then discusses IIB, both the case of O3/O7 orientifold models in section~\ref{O3O7Case}, and of O5/O9 models in section~\ref{O5O9Case}.  In particular, in all of these cases we work out explicit expressions for the superpotential and the D-terms, which, when combined with the K\"ahler potential and holomorphic gauge kinetic couplings computed in~\cite{Grimm:2004ua,Grimm:2004uq}, completely specifies the effective theory.  In section~\ref{Examples} we construct some IIB examples which exhibit D-terms of the sort we describe, and we explain why such examples are slightly tricky to come by.  Finally, in section~\ref{Conclusions} we summarize our results, and mention some open problems and directions for future work.

\section{Generalized NS-NS Fluxes and D-Terms in IIA}
\label{IIACase}

In this section we will review the results of~\cite{Ihl:2007ah}.

\subsection{IIA orientifolds with the usual fluxes}

Let us first establish some conventions for the IIA orientifolds that we will be discussing.  Let $X$ be a Calabi-Yau three-fold, and let $\s$ be an anti-holomorphic involution of $X$.  The cohomology of $X$ then splits into even and odd parts, depending upon the behavior of each class under $\s$.  We will take the following basis of representative forms:
\begin{itemize}
\item The zero-form 1,
\item a set of odd two-forms $\om_a$, $a=1,\ldots,h^{1,1}_-$,
\item a set of even two-forms $\m_\al$, $\al=1,\ldots,h^{1,1}_+$,
\item a set of even four-forms $\widetilde\om^a$, $a=1,\ldots,h^{1,1}_-$,
\item a set of odd four-forms $\widetilde\m^\al$, $\al=1,\ldots,h^{1,1}_+$,
\item a six form $\varphi$, odd under $\s$,
\item a set of even three-forms $a_K$, $K=1,\ldots,h^{2,1}+1$,
\item and a set of odd three-forms $b^K$, $K=1,\ldots,h^{2,1}+1$.
\end{itemize}
Additionally, it turns out that we can always choose the $a_K$ and $b^K$ to form a symplectic basis such that the only non-vanishing intersections are
\be
\int_X a_K\w b^J=\d_K^J.
\ee

For the even-degree forms we will allow ourselves a bit more freedom of scaling, in order to simplify some explicit computations in the case of toroidal orientifold examples.  We will take the intersections to be
\be
\int_X\varphi = f,\qquad\int_X\om_a\w\om_b\w\om_c=\k_{abc},\qquad\int_X\om_a\w\m_\al\w\m_\beta=\widehat{\k}_{a\,\al\beta},\non
\ee
\be
\int_X\om_a\w\widetilde{\om}^b=d_a\vphantom{d_a}^b,\qquad\int_X\m_\al\w\widetilde{\m}^\beta=\widehat{d}_\al\vphantom{\widehat{d}}^\beta.
\ee
If we chose the four-forms to be a basis dual to the two forms, then we would of course set $d_a\vphantom{d_a}^b=\d_a^b$, $\widehat{d}_\al\vphantom{\widehat{d}}^\beta=\d_\al^\beta$, but we will prefer instead to leave things here more general\footnote{Note however that Poincar\'e duality implies in this case that $d$ and $\widehat d$ are both invertible matrices.  Indeed we will need to use this fact to write explicit expressions below.}.

Now let us describe the four-dimensional fields of this class of compactifications, restricting ourselves, for simplicity, to the bosonic sector.  First we have the K\"ahler moduli, parametrized by complex scalar fields $t^a=u^a+iv^a$ coming from the expansion
\be
B+iJ=J_c=t^a\om_a,
\ee
where the complexified K\"ahler form $J_c$ must be odd under $\s$.  Note that the K\"ahler form $J=v^a\om_a$ determines the compactification volume (in string frame) via
\be
\mathcal{V}_6=\frac{1}{3!}\int_XJ\w J\w J=\frac{1}{6}\k_{abc}v^av^bv^c.
\ee

To describe the complex moduli, let us write the holomorphic three-form as
\be
\Om=\mathcal{Z}^Ka_K-\mathcal{F}_Kb^K.
\ee
We will use conventions in which
\be
i\int_X\Om\w\bar\Om=1,\qquad\s^\ast\Om=\bar\Om,
\ee
so that the $\mathcal{Z}^K$ are real functions of the complex moduli and $\mathcal{F}_K$ are pure imaginary, and together they satisfy the constraint $\mathcal{Z}^K\mathcal{F}_K=-i/2$.  We can now define a complexified version~\cite{Grimm:2004ua}
\be
\Om_c=C_3+2ie^{-D}\Re\Om=\lp\xi^K+2ie^{-D}\mathcal{Z}^K\rp a_K,
\ee
where $e^{-D}=\mathcal{V}_6^{1/2}e^{-\phi}$ contains the dilaton and we expand the periods of $C_3$ (which must be even under $\s$ in order to survive the orientifold projection) as $C_3=\xi^Ka_K$.  Note that we abuse notation somewhat here as we ignore other pieces which contribute to the ten-dimensional R-R three-form potential $C_3$, namely pieces that give rise to four-dimensional vectors and (local) pieces that give the four-form R-R flux, both of which will be discussed below.  The complex moduli $N^K=\hlf\xi^K+ie^{-D}\mathcal{Z}^K$ are then simply given by the expansion
\be
\Om_c=2N^Ka_K,
\ee
and include the complex structure moduli of the metric, the dilaton, and the R-R three-form periods.

Next we turn to the four-dimensional vectors that come from reducing $C_3$ against the forms $\m_\al$, so that the total field $C_3$ (before turning on fluxes) is
\be
C_3=\xi^Ka_K+A^\al\w\m_\al,
\ee
with the $A^\al$ being one-form gauge potentials in four-dimensions.  We will associate these potentials to electric $\U(1)$ gauge groups in the four-dimensional effective theory, but we will also later be interested in the dual magnetic $\U(1)$s.  These are associated to dual one-forms obtained by reducing $C_5$ against odd four-forms,
\be
C_5=\widetilde A_\al\w\widetilde\m^\al.
\ee
Note that there are no vectors arising from $C_1$ or $C_7$, because these are projected out by the orientifold.

These account for our bosonic fields in four dimensions.  We would also like to include fluxes from R-R field strengths and from the NS-NS field strength $H$ (we will include more general NS-NS fluxes below).  Expanding in our cohomological basis, we have
\be
F_0=m_0,\qquad F_2=m^a\om_a,\qquad F_4=e_a\widetilde\om^a,\qquad F_6=e_0\varphi,
\ee
and
\be
H=p_Kb^K.
\ee

Another crucial point to keep in mind is that the ten-dimensional action includes a piece\footnote{The unusual factors of $\sqrt 2$ are an unfortunate consequence of our normalizations, which follow~\cite{Grimm:2004ua,DeWolfe:2005uu}. Note also that we have set $\al'=4 \pi^2$.}
\be
\label{C7Tadpole}
\int_{\R^4\times X}\left\{-\hlf\lp F_2+m_0B_2\rp\w\ast\lp F_2+m_0B_2\rp+C_7\w\ls\frac{1}{\sqrt 2}\d_{D6}-\sqrt 2\d_{O6}\rs\right\}.
\ee
Since $\ast(F_2+m_0B_2)=dC_7+\cdots$, the vanishing of the $C_7$ tadpole then implies the constraint
\be
-m_0p_Kb^K+\frac{1}{\sqrt 2}\ls\d_{D6}\rs=\sqrt 2\ls\d_{O6}\rs,
\ee
(though note that the tadpole condition is actually stronger than this cohomological constraint).

If we are in a regime where a four-dimensional effective description is expected to be valid, then it is useful to assemble the data just described into an $\N=1$ four-dimensional effective supergravity theory.  Such a theory consists of one gravity multiplet, some number of chiral multiplets, including complex scalars $\phi^I$, and some number of vector multiplets including vectors $A^\al$.  The theory is then specified by giving three functions which will depend on the complex scalars, namely a K\"ahler potential $K$, a holomorphic superpotential $W$, and holomorphic gauge-kinetic couplings $f_{\al\beta}$.  The bosonic part of the effective action is then
\bea
\label{BosonicAction}
S^{(4)} &=& -\int_{M_4}\left\{-\hlf R\ast 1+K_{I\bar J}d\phi^I\w\ast d\bar\phi^{\bar J}+V\ast 1\right.\non\\
&& \left.+\hlf\lp\Re f_{\al\beta}\rp F^\al\w\ast F^\beta+\hlf\lp\Im f_{\al\beta}\rp F^\al\w F^\beta\right\},
\eea
where the scalar potential is
\be
V=e^K\lp K^{I\bar J}D_IW\overline{D_JW}-3|W|^2\rp+\hlf\lp\Re f\rp^{-1\,\al\beta}D_\al D_\beta.
\ee
Here, $\ast$ is the four-dimensional Hodge star, $K_{I\bar J}=\p_I\bar\p_{\bar J}K$, $K^{I\bar J}$ is its (transpose) inverse, $F^\al=dA^\al$, and $D_IW=\p_IW+(\p_IK)W$.  $D_\al$ is the D-term for the $\U(1)$ gauge group corresponding to $A^\al$, which in four dimensional $\mathcal{N}=1$ SUGRA is given by~\cite{Binetruy:2004hh,Choi:2005ge} (for field configurations with $W\ne 0$)
\be
\label{GeneralDTerm}
D_\al=\frac{i}{W}\d_\al\phi^ID_IW=i\p_IK\d_\al \phi^I+i\frac{\d_\al W}{W},
\ee
where $\lambda^\al\d_\al\phi^I$ is the variation of the field $\phi^I$ under an infinitesimal gauge transformation $A^\al\rightarrow A^\al+d\lambda^\al$.  The second term above, proportional to the gauge variation of the superpotential, is to be interpreted as a Fayet-Iliopoulos term.  It occurs, for instance, when we have gauged an R-symmetry.  It will turn out that in our constructions, the superpotential will always remain gauge neutral, and hence we will not generate any F-I terms, and we will always be able to write (even if $W=0$)
\be
\label{SpecialDterm}
D_\al=i\p_IK\d_\al\phi^I.
\ee

Now we plug in the fields and fluxes above into the ten-dimensional SUGRA action, perform a Ka{\l}u\.za-Klein reduction to four dimensions, and compare to the action (\ref{BosonicAction}), following~\cite{Grimm:2004ua}.  From the kinetic terms we find
\be
f_{\al\beta}=i\widehat\k_{a\,\al\beta}t^a,
\ee
and
\be
K=4D-\ln\lp\frac{4}{3}\k_{abc}v^av^bv^c\rp.
\ee
From the potential terms we then find that
\be
\label{IIABasicSuperpotential}
W=\int_X\Om_c\w H+\int_Xe^{J_c}\w F_{RR}=2N^Kp_K+fe_0+d_a\vphantom{d_a}^be_bt^a+\hlf\k_{abc}m^at^bt^c+\frac{1}{6}m_0\k_{abc}t^at^bt^c.
\ee
Here $F_{RR}=F_0+F_2+F_4+F_6$ is the formal sum of R-R fluxes, and
\be
e^{J_c}=1+J_c+\hlf J_c\w J_c+\frac{1}{6}J_c\w J_c\w J_c.
\ee
Also, the D-terms in this setup vanish, $D_\al=0$.

\subsection{Metric fluxes}

Let us restrict for the moment to the case of toroidal orientifolds.  It is well known that by T-dualizing one circle of a torus with $H$-flux, one can swap some components of the $H$-flux for some non-constant metric components.  The new geometry that results is called a twisted torus, and the one-forms $dx^i$ are no longer globally defined.  Instead, they should be replaced by one-forms $\eta^i$ which are globally defined\footnote{In fact, all of $\Om^\ast(X)$, where $X$ is the twisted torus, is generated by wedge products of the $\eta^i$ with coefficients being globally defined functions.}, but which are no longer necessarily closed, satisfying instead
\be
\label{detai}
d\eta^i=-\hlf\om^i_{jk}\eta^j\w\eta^k,
\ee
where $\om^i_{jk}$ are constant coefficients, antisymmetric in the lower two indices.  These coefficients are known as metric (or sometimes geometric) fluxes, and arise, like $H$-flux, from the NS-NS sector of the theory.

By taking the exterior derivative of (\ref{detai}), we find a consistency condition
\be
\label{omomBianchi}
\om^m_{[ij}\om^n_{k]m}=0,\qquad\forall n,i,j,k.
\ee
In fact, rather than proceeding by T-duality, we can take (\ref{detai}) and (\ref{omomBianchi}) as a starting point for defining a twisted torus $X$~\cite{Scherk:1979zr,Kaloper:1999yr}.  We will also impose the additional constraint of tracelessness,
\be
\om^i_{ij}=0,\qquad\forall j,
\ee
but we will occasionally point out how relaxing this condition would modify our results (relaxing this condition would for example have the effect that the na\"ive volume form of the twisted torus would be exact~\cite{Wecht:2007wu}, but it is not immediately obvious that this is contradictory).

It is natural to consider also $H$-flux on $X$, which should be a globally defined three-form
\be
H=\frac{1}{6}H_{ijk}\eta^i\w\eta^j\w\eta^k,
\ee
and must still be closed, leading to the identity
\be
\label{omHBianchi}
\om^i_{[jk}H_{\ell m]i}=0.
\ee
We are assuming here that the coefficients $H_{ijk}$ are constant.  In some specific twisted torus cases it can be checked explicitly that each cohomology class has a representative with this property (i.e. constant coefficients in an $\eta^i$ expansion), and we believe that this will hold in general.  Together, (\ref{omomBianchi}) and (\ref{omHBianchi}) are known as Bianchi identities.

On a toroidal orientifold we should have both $H_{ijk}$ and $\om^i_{jk}$ invariant under the orbifold group.  Under the involution $\s$ we should have $\s^\ast H=-H$, since $B_2$ is odd under world-sheet parity $\Om_p$.  The metric is even under $\Om_p$ and hence should be invariant under $\s$, and since the $\om^i_{jk}$ essentially appear as coefficients in the metric, they too should be even under $\s$ (for a more convincing explanation see~\cite{Ihl:2007ah}, or references therein).

Recall that if we applied our discussion of the four-dimensional effective theory above to the toroidal case, then it makes much more sense to describe $H$-flux not in terms of components $H_{ijk}$, but rather by coefficients $p_K$, i.e. $H=p_Kb^K$, where $b^K$ are a basis for the odd untwisted three-forms of the toroidal orientifold.  A similar choice is convenient for the metric fluxes.  Consider a general $p$-form
\be
A=\frac{1}{p!}A_{i_1\cdots i_p}\eta^{i_1}\w\ldots\w\eta^{i_p}.
\ee
Let's assume for now that the $A_{i_1\cdots i_p}$ are constants.  In that case, we can define a $(p+1)$-form $\om\cdot A=-dA$, which in components reads
\be
\label{omcdotA}
\lp\om\cdot A\rp_{i_1\cdots i_{p+1}}=\binom{p+1}{2}\om^j_{[i_1i_2}A_{|j|i_3\cdots i_{p+1}]},
\ee
and where we use the convention that $\binom{n}{m}=0$ unless $0\le m\le n$.

As a brief aside, note that we can take (\ref{omcdotA}) as a definition of the $(p+1)$-form $\om\cdot A$ even when the original components $A_{i_1\cdots i_p}$ are not constant\footnote{The appropriate generalization when $\om^i_{jk}$ are not traceless is
\be
\lp\om\cdot A\rp_{i_1\cdots i_{p+1}}=\binom{p+1}{2}\om^j_{[i_1i_2}A_{|j|i_3\cdots i_{p+1}]}+\hlf\binom{p+1}{1}\om^j_{j[i_1}A_{i_2\cdots i_{p+1}]}.\non
\ee
}.  In this case we can write $dA=d'A-\om\cdot A$, where $d'$ is understood to act only on the coefficients $A_{i_1\cdots i_P}$.  This then inspires an approach that will be useful later when we will add non-geometric fluxes as well.  Rather than work on the twisted torus with forms expanded in $\eta^i$ and exterior derivative $d$, we can work on the flat torus with forms $dx^i$ and replace the exterior derivative by
\be
d_\om=d-\om\cdot.
\ee
In fact, in the presence of $H$-flux, the natural derivative acting on R-R forms is $d_H=d+H\w$.  In the language above we can either work with the twisted torus and forms $\eta^i$, with derivative $d_H$, where $H$ is also expanded in the $\eta^i$, or we can work on the flat torus with forms $dx^i$ and exterior derivative
\be
d_{H,\om}=d+H\w-\om\cdot.
\ee
This latter approach will be the one which naturally generalizes to the ``non-geometric'' case.  Note that the requirement $d_{H,\om}^2=0$ reproduces both of our Bianchi identities above.

Taking either of these two perspectives, we are now ready to define a cohomological parametrization for the metric fluxes, in analogy with the $p_K$.  We simply take a basis for the untwisted two-forms of the toroidal orientifold, with $\om_a$ being odd and $\m_\al$ being even.  Then we expand
\be
\om\cdot\om_a=r_{aK}b^K,\qquad\om\cdot\m_\al=\widehat{r}_\al^Ka_K.
\ee

Integration by parts then also furnishes the expansions
\be
\om\cdot a_K=\lp d^{-1}\rp_a\vphantom{\lp d^{-1}\rp}^br_{bK}\widetilde{\om}^a,\qquad\om\cdot b^K=-\lp\widehat{d}^{-1}\rp_\al\vphantom{\lp\widehat{d}^{-1}\rp}^\beta\widehat{r}_\beta^K\widetilde{\m}^\al.
\ee
These coefficients $r_{aK}$ and $\widehat{r}_\al^K$ are the analogues of the $p_K$.  Indeed, in the case of $H$-flux the corresponding expressions would be
\be
H\w 1=p_Kb^K,\qquad H\w a_K=-f^{-1}p_K\varphi.
\ee
The great promise of these cohomological parametrizations of the NS-NS fluxes is that they can be generalized beyond the toroidal case; since the $p_K$, $r_{aK}$ and $\widehat{r}_\al^K$ are defined only in terms of maps between representatives of the untwisted cohomology of the toroidal orbifold, we can try to define similar maps between cohomological representatives on any Calabi-Yau space which admits an orientifold involution.  In the case of $p_K$, this is of course completely standard for parametrizing possible $H$-flux.  In general the matrices $r_{aK}$ and $\widehat{r}_\al^K$ will be $h^{1,1}_-\times(h^{2,1}+1)$ and $h^{1,1}_+\times(h^{2,1}+1)$ matrices, respectively.

By requiring $d_{H,\om}^2$ to vanish on the invariant forms, we learn that the Bianchi identities imply some relations among these coefficients.  In particular,
\be
\label{CohomologicalMetricBianchis}
p_K\widehat{r}_\al^K=0,\quad\forall\al,\qquad r_{aK}\widehat{r}_\al^K=0,\quad\forall a,\al.
\ee
Unfortunately, it turns out that these are not the complete set of Bianchi identities; the requirement that $d_{H,\om}^2=0$ also on non-invariant forms is stronger.  This is especially vexing in that it is not clear what these extra Bianchi constraints should be once one moves beyond toroidal examples.

There is one more caveat worth noting in this approach.  For $H$-flux it is automatically true that the odd invariant combinations of flux components $H_{ijk}$ are in a bijective correspondence with the odd invariant untwisted three-forms, so the $p_K$ really do describe all the possible $H$-fluxes we would like to turn on.  This is no longer the case with the metric fluxes; it is not necessarily true that the number of invariant combinations of $\om^i_{jk}$ is equal to the number of $r_{aK}$ and $\widehat{r}_\al^K$.  The count of the latter coefficients is given by $\hlf b^2b^3=h^{1,1}(h^{2,1}+1)$, with the Betti and Hodge numbers here referring to the untwisted sector of the orbifold.  In many examples the bijective correspondence does hold.  For instance in orientifolds built from orbifolds $T^6/\G$, where $\G$ is any of the crystallographic actions $\Z_2\times\Z_2$, $\Z_3$, $\Z_3\times\Z_3$ (as in, e.g.~\cite{DeWolfe:2005uu}), $\Z_4$ (as in~\cite{Ihl:2007ah}), or $\Z_{6-I}$, the cohomological parameters capture all of the possible metric fluxes.  Note that the nature of the involution here is irrelevant for the counting.  However, in some other examples, like $\Z_{6-II}$, there are more possible combinations of $\om^i_{jk}$ than there are components of $r_{aK}$ and $\widehat{r}_\al^K$ (in th $\Z_{6-II}$ case there are seven invariant combinations of metric flux, but only $\hlf 3\cdot 4=6$ cohomological parameters).

These ``extra'' fluxes do not, however, seem to appear in the four-dimensional effective action.  The metric fluxes will contribute to the superpotential only through $r_{aK}$, and contribute to the D-terms only through $\widehat{r}_\al^K$.

Observe that the fluxes $F_2=m^a\om_a$ are no longer closed.  Looking at (\ref{C7Tadpole}) we see that this results in a new contribution to the $C_7$ tadpole,
\be
-\sqrt 2\lp m_0p_K-m^ar_{aK}\rp b^K+\ls\d_{D6}\rs=2\ls\d_{O6}\rs.
\ee
Actually, this is most naturally expressed by noting that the flux contributions to the tadpole are naturally proportional to
\be
d_{H,\om}F_{RR}|_{\mathrm{3-form}}=H F_0-\om\cdot F_2.
\ee

There are two avenues towards understanding the effect of these metric fluxes on the four-dimensional effective theory.  One can either use T-duality to deduce the way in which the metric fluxes appear in quantities like the superpotential~\cite{Camara:2005dc}, or one can explicitly perform a Ka{\l}u{\.z}a-Klein reduction on a twisted torus~\cite{Villadoro:2005cu}.  Either method will reveal that the addition of metric fluxes has two effects on the four-dimensional effective theory.  First of all, the superpotential (\ref{IIABasicSuperpotential}) gets modified by the addition of a term $2N^Kr_{aK}t^a$, so that it can be written
\bea
\label{MetricFluxSuperpotential}
W &=& \int_X\Om_c\w d_{H,\om}\lp e^{-J_c}\rp+\int_Xe^{J_c}\w F_{RR}\\
&=& 2N^K\lp p_K+r_{aK}t^a\rp+fe_0+d_a\vphantom{d_a}^be_bt^a+\hlf\k_{abc}m^at^bt^c+\frac{1}{6}m_0\k_{abc}t^at^bt^c.\non
\eea

The second effect is to charge some of the moduli under the electric gauge groups $\U(1)^\al$.  Indeed, recall that the gauge vectors descended from the three-form potential which had the expansion
\be
C_3=A^\al\w\m_\al+\xi^Ka_K,
\ee
where we ignore the (local) parts of $C_3$ which contribute to the four-form flux $e_a\widetilde\om^a$.  In the case without metric fluxes, the four-dimensional gauge transformations $A^\al\rr A^\al+d\lambda^\al$ are the descendants of the ten-dimensional three-form gauge transformations
\be
\label{IIAElectricGaugeTransformation}
C_3\longrightarrow C_3+d\lp\lambda^\al\m_\al\rp=\lp A^\al+d\lambda^\al\rp\w\m_\al+\xi^Ka_K.
\ee
In particular, this ten-dimensional transformation can be done without modifying any of the four-dimensional fields; all the scalars are neutral under these gauge groups.

However, in the presence of metric fluxes $\widehat{r}_\al^K$, $\m_\al$ is no longer closed, and the transformation above gets modified to
\be
C_3\longrightarrow C_3+d\lp\lambda^\al\m_\al\rp=\lp A^\al+d\lambda^\al\rp\w\m_\al+\lp\xi^K-\lambda^\al\widehat{r}_\al^K\rp a_K,
\ee
and we see that the four-dimensional field $\xi_K$ is no longer left invariant under this electric gauge transformation.

To calculate the resulting D-terms from (\ref{SpecialDterm}) we need one relation, namely that
\be
\frac{\p}{\p N^K}D=-e^D\mathcal{F}_K.
\ee
We can then compute
\be
D_\al=2ie^D\widehat{r}_\al^K\mathcal{F}_K.
\ee
Recall that $\mathcal{F}_K$ in our conventions is pure imaginary, so that $D_\al$ is real.

Thus, in a supersymmetric vacuum, we will have to solve not only the F-term equations from the superpotential (\ref{MetricFluxSuperpotential}), but also the D-term equations,
\be
D_\al=0\qquad\Longrightarrow\qquad\widehat{r}_\al^K\mathcal{F}_K=0.
\ee
If we are in a non-supersymmetric vacuum, then the scalar potential now has a D-term piece,
\be
V_D=\hlf\lp\Re f\rp^{-1\,\al\beta}D_\al D_\beta=2e^{2D}\lp\widehat\k v\rp^{-1\,\al\beta}\lp\mathcal{F}_K\widehat{r}_\al^K\rp\lp\mathcal{F}_J\widehat{r}_\beta^J\rp,
\ee
where we have used
\be
\Re f_{\al\beta}=-\widehat\k_{a\,\al\beta}v^a=-\lp\widehat\k v\rp_{\al\beta}.
\ee

Finally, we verify that the F-I terms are zero by calculating the variation of the superpotential $W$ under gauge transformations.  Indeed, we find that under (\ref{IIAElectricGaugeTransformation}) we have
\be
\d W=-\lambda^\al\widehat{r}_\al^K\lp p_K+r_{aK}t^a\rp=0,
\ee
where we have used the cohomological Bianchi identities (\ref{CohomologicalMetricBianchis}).

\subsection{Non-geometric fluxes}

By T-dualizing two legs of $H$-flux on a torus, the usual Buscher rules~\cite{Buscher:1987sk} lead one to construct a background that is no longer a globally defined geometry (though one can interpret it as a locally geometric toroidal fiber over a geometric base).  The parameters describing this construction are components $Q^{ij}_k$, antisymmetric in the upper indices, and are analogous to $H_{ijk}$ and $\om^i_{jk}$.  At the level of the effective four-dimensional theory, it is natural to also introduce the totally antisymmetric $R^{ijk}$, which one can formally imagine as resulting from T-dualizing all three legs of toroidal $H$-flux~\cite{Shelton:2005cf}.  From a ten-dimensional perspective, it is not clear how to construct such a thing (which would not admit even a local geometric interpretation~\cite{Shelton:2006fd}) since the need to choose an initial trivialization for the $H$-flux breaks one of the three necessary isometries\footnote{In fact, there are also examples of $Q$-fluxes and metric fluxes $\om^i_{jk}$ which seem very difficult to construct from a ten-dimensional viewpoint.  For a subset which can be constructed, see~\cite{Ihl:2007ah}.}.  However, it does not inconvenience us in our current context to include all these possible non-geometric fluxes, so we shall. Like $H_{ijk}$ and $\om^i_{jk}$, they all arise from the NS-NS sector.

Under an orientifold involution, the $Q$-fluxes should be odd, like $H$-flux, while the $R$-fluxes should be even, like metric flux (one justification for this is that a pair of T-dualities should preserve the world-sheet parity eigenvalue).  As with our previous examples, it is natural to define actions of these fluxes on (components of) differential forms.  First, the $Q$-fluxes allow a map from $p$-forms to $(p-1)$-forms\footnote{This is actually assuming a tracelessness condition, $Q^{jk}_j=0$.  If this condition is dropped, then the correct generalization would be
\be
\lp Q\cdot A\rp_{i_1\cdots i_{p-1}}=\hlf\binom{p-1}{1}Q^{jk}_{[i_1}A_{|jk|i_2\cdots i_{p-1}]}+\hlf\binom{p-1}{0}Q^{jk}_jA_{ki_1\cdots i_{p-1}}.\non
\ee
},
\be
\label{QAction}
\lp Q\cdot A\rp_{i_1\cdots i_{p-1}}=\hlf\binom{p-1}{1}Q^{jk}_{[i_1}A_{|jk|i_2\cdots i_{p-1}]},
\ee
while the $R$-fluxes map $p$-forms to $(p-3)$-forms,
\be
\label{RAction}
\lp R\cdot A\rp_{i_1\cdots i_{p-3}}=\frac{1}{6}\binom{p-3}{0}R^{jk\ell}A_{jk\ell i_1\cdots i_{p-3}}.
\ee
The inclusion of the somewhat trivial binomial coefficients is simply to make it clear that $Q$ kills forms below degree two, while $R$ kills forms below degree three.

With these actions, it is convenient to define a differential operator which acts on forms~\cite{Shelton:2006fd,Micu:2007rd},
\be
\label{IIAD}
\mathcal{D}=d+H\w-\om\cdot+Q\cdot-R\cdot.
\ee
Requiring that $\mathcal{D}^2=0$ leads to a set of Bianchi identities,
\bea
\label{IIANongeometricBianchis}
H_{m[ij}\om^m_{k\ell]} &=& 0,\non\\
H_{m[ij}Q^{m\ell}_{k]}-\om^m_{[ij}\om^\ell_{k]m} &=& 0,\non\\
H_{ijm}R^{k\ell m}+\om^m_{ij}Q^{k\ell}_m-4\om^{[k}_{m[i}Q^{\ell]m}_{j]} &=& 0,\\
\om^{[j}_{mi}R^{k\ell]m}-Q^{[jk}_mQ^{\ell]m}_i &=& 0,\non\\
Q^{[ij}_mR^{k\ell]m} &=& 0,\non
\eea
along with the additional requirement that $H_{ijk}R^{ijk}=\om^i_{jk}Q^{jk}_i=0$, which is trivially satisfied on orientifolds since there are no odd invariant scalars.

It is again natural to introduce cohomological parameters via the expansions
\be
Q\cdot\widetilde\om^a=q^a_Kb^K,\qquad Q\cdot\widetilde\m^\al=\widehat{q}^{\al K}a_K,
\ee
\be
R\cdot\varphi=s_Kb^K.
\ee
We then have also
\be
Q\cdot a_K=-\lp d^{-1}\rp_a\vphantom{\lp d^{-1}\rp}^bq^a_K\om_b,\qquad Q\cdot b^K=\lp\widehat{d}^{-1}\rp_\al\vphantom{\lp\widehat{d}^{-1}\rp}^\beta\widehat{q}^{\al K}\m_\beta,
\ee
\be
R\cdot a_K=f^{-1}s_K1.
\ee

Again, some, but not all, of the Bianchi identities follow from demanding that $\mathcal{D}^2$ vanish on our cohomological basis, namely
\bea
\widehat{r}_\al^Kp_K=\widehat{r}_\al^Ks_K=\widehat{q}^{\al K}p_K=\widehat{q}^{\al K}s_K &=& 0,\quad\forall\al,\non\\
\label{IIANongeometricCohomologicalBianchis}
\widehat{r}_\al^Kr_{bK}=\widehat{r}_\al^Kq^b_K=\widehat{q}^{\al K}r_{bK}=\widehat{q}^{\al K}q^b_K &=& 0,\quad\forall\al,b,\\
f^{-1}p_{[K}s_{J]}+\lp d^{-1}\rp_a\vphantom{\lp d^{-1}\rp}^br_{b[K}q^a_{J]}=\lp\widehat{d}^{-1}\rp_\al\vphantom{\lp\widehat{d}^{-1}\rp}^\beta\widehat{q}^{\al[K}\widehat{r}_\beta^{J]} &=& 0,\quad\forall K,J.\non
\eea

Also, we still have the possibility that there is not a bijective mapping between flux parameters $Q^{ij}_k$ and cohomological parameters $q^a_K$ and $\widehat{q}^{\al K}$, but the same remarks apply as for metric fluxes.  There is always a bijective correspondence for the $R$-fluxes.

The modifications to the tadpole condition can be obtained by T-duality arguments,
\be
\label{IIAGeneralTadpole}
-\sqrt 2\mathcal{D}F_{RR}+\ls\d_{D6}\rs=2\ls\d_{O6}\rs,
\ee
or
\be
-\sqrt 2\lp p_Km_0-r_{aK}m^a+q^a_Ke_a-s_Ke_0\rp+N_K^{(D6)}=2N_K^{(O6)}.
\ee

Similar arguments can also be used to obtain the superpotential~\cite{Aldazabal:2006up,Micu:2007rd},
\bea
W &=& \int_Xe^{J_c}\w F_{RR}+\int_X\Om_c\w\mathcal{D}\lp e^{-J_c}\rp\non\\
&=& fe_0+d_a\vphantom{d}^bt^ae_b+\hlf\k_{abc}t^at^bm^c+\frac{1}{6}m_0\k_{abc}t^at^bt^c\non\\
&& +2N^K\lp p_K+r_{aK}t^a+\hlf\k_{abc}\lp d^{-1}\rp_d\vphantom{\lp d^{-1}\rp}^aq^d_Kt^bt^c+\frac{1}{6}f^{-1}s_K\k_{abc}t^at^bt^c\rp.
\eea

Now using this newly defined exterior derivative $\mathcal{D}$, we propose that the proper R-R gauge transformations should be encoded as
\be
C_{RR}\longrightarrow C_{RR}+\mathcal{D}\Lambda,
\ee
where $\Lambda$ is a formal sum of even forms.  Explicitly, if $\Lambda$ is a four-dimensional scalar times a set of internal forms, then the orientifold projection actually forces
\be
\Lambda=\lambda^\al\m_\al+\widetilde\lambda_\al\widetilde\m^\al,\qquad\mathcal{D}\Lambda=d\lambda^\al\w\m_\al+d\widetilde\lambda_\al\w\widetilde\m^\al+\lp\widehat{q}^{\al K}\widetilde\lambda_\al-\widehat{r}_\al^K\lambda^\al\rp a_K.
\ee
From this we can see our earlier conclusion that $\lambda^\al$ generates gauge transformations in the electric gauge groups $\U(1)^\al$, and that the $\widehat{r}_\al^K$ correspond to electric charges.  But we also see that $\widetilde\lambda_\al$ generates gauge transformations in the corresponding {\it{magnetic}} gauge groups $\widetilde{\U(1)}_\al$ (whose vectors come from $C_5$ reduced against $\widetilde\m^\al$) and that the non-geometric fluxes $\widehat{q}^{\al K}$ correspond to magnetic charges.

We should again quickly check whether the superpotential remains neutral under these magnetic gauge transformations as claimed above.  Indeed,
\be
\d W=\widetilde\lambda_\al\widehat{q}^{\al K}\lp p_K+r_{aK}t^a+\hlf\k_{abc}\lp d^{-1}\rp_d\vphantom{\lp d^{-1}\rp}^aq^d_Kt^bt^c+\frac{1}{6}f^{-1}s_K\k_{abc}t^at^bt^c\rp=0,
\ee
where we have used the first two lines of (\ref{IIANongeometricCohomologicalBianchis}), so there are no F-I terms generated.

Since we have now electric and magnetic charges and dyonic fields (i.e. carrying potentially both electric and magnetic charges), it is interesting to ask if the collection of charged scalars remains mutually local.  The condition that two charged scalars be mutually local is
\be
\label{IIAMutuallyLocal}
\lp\widehat{d}^{-1}\rp_\al\vphantom{\lp\widehat{d}^{-1}\rp}^\beta\lp\widehat{q}^{\al K}\widehat{r}_\beta^J-\widehat{q}^{\al J}\widehat{r}_\beta^K\rp=0.
\ee
Note that under the usual normalization for the dual gauge groups, the fields $\widetilde{A}_\al$ should be rescaled by the matrix $\widehat{d}_\beta\vphantom{\widehat{d}}^\al$, or equivalently the magnetic charges $\widehat{q}^{\al K}$ should be rescaled by $\widehat{d}^{-1}$,
so that in our conventions the correct mutual locality condition is as above.  But (\ref{IIAMutuallyLocal}) is simply the final equation of (\ref{IIANongeometricCohomologicalBianchis}) and thus is guaranteed by the Bianchi identities.

The mutual locality in turn implies that there always exists a $\Sp(2h^{1,1}_+;\Z)$ transformation which can rotate all the charges to be electric charges\footnote{This statement actually relies also on charge quantization, which we have not demonstrated here.  A more detailed discussion of the subtleties can be found in section \ref{IIBCase}.}.  The resulting electric gauge groups after rotation will have associated D-terms which must vanish in any supersymmetric solution.  However, for the moment it will be more convenient to use our original basis of gauge groups, but include also magnetic contributions.  Recall that the holomorphic gauge kinetic couplings for the electric groups were given by
\be
f_{\al\beta}=i\lp\widehat\k t\rp_{\al\beta}.
\ee
Similar calculations (by reducing the piece of the ten-dimensional action which is quadratic in $C_5$) give the holomorphic magnetic gauge kinetic couplings,
\be
\widetilde f^{\al\beta}=-i\lp\widehat\k t\rp^{-1\,\g\d}\widehat{d}_\g\vphantom{\widehat{d}}^\al\widehat{d}_\d\vphantom{\widehat{d}}^\beta.
\ee
The magnetic analogs of our previous electric D-terms are
\be
\widetilde{D}^\al=-2ie^D\widehat{q}^{\al K}\mathcal{F}_K,
\ee
and the resulting D-term contribution to the scalar potential is
\bea
V_D &=& \hlf\lp\Re f\rp^{-1\,\al\beta}D_\al D_\beta+\hlf\lp\Re\widetilde f\rp^{-1}_{\al\beta}\widetilde{D}^\al\widetilde{D}^\beta\non\\
&=& -2e^{2D}\ls\lp\Re f\rp^{-1\,\al\beta}\widehat{r}_\al^K\widehat{r}_\beta^J+\lp\Re\widetilde f\rp^{-1}_{\al\beta}\widehat{q}^{\al K}\widehat{q}^{\beta J}\rs\mathcal{F}_K\mathcal{F}_J.
\eea

Though not immediately apparent in this form, this expression is positive semi-definite (the gauge kinetic couplings are positive definite and the D-terms are real), and must vanish in a supersymmetric vacuum.  Note that this piece of the potential can have reasonably complicated dependence on all of the scalar fields.

\section{Generalized NS-NS Fluxes and D-Terms in IIB}
\label{IIBCase}

We will now follow a very similar procedure in the case of IIB.  In the context of IIB, a Calabi-Yau orientifold which doesn't explicitly break supersymmetry (though the inclusion of fluxes will allow for spontaneous breaking) must be paired with an involution $\s$ which is a holomorphic isometry of the Calabi-Yau three-fold $X$.  It turns out that this still leaves two broad classes of orientifolds, differentiated by their action on the holomorphic $(3,0)$-form $\Om$.  If $\s^\ast\Om=-\Om$, then the fixed point set of $\s$ can have complex codimension three or one, so we call these O3/O7 orientifolds, while if $\s^\ast\Om=\Om$, then the codimension is two or zero, and we refer to O5/O9 orientifolds.  The full orientifold $\Z_2$ action is generated by $(-1)^{F_L}\Om_p\s$ in the former case and $\Om_p\s$ in the latter case, where $F_L$ is the spacetime fermion number in the left-moving sector, and $\Om_p$ is the world-sheet parity operator.

We will treat the two cases separately, but we can use a common cohomological basis.  In even degree we have
\begin{itemize}
\item The zero-form 1,
\item a set of even two-forms $\m_\al$, $\al=1,\ldots,h^{1,1}_+$,
\item a set of odd two-forms $\om_a$, $a=1,\ldots,h^{1,1}_-$,
\item a set of even four-forms $\widetilde\m^\al$, $\al=1,\ldots,h^{1,1}_+$,
\item a set of odd four-forms $\widetilde\om^a$, $a=1,\ldots,h^{1,1}_-$,
\item a six form $\varphi$, even under $\s$,
\end{itemize}
with intersections
\be
\int_X\varphi = f,\qquad\int_X\m_\al\w\m_\beta\w\m_\g=\k_{\al\beta\g},\qquad\int_X\m_\al\w\om_a\w\om_b=\widehat\k_{\al\,ab},\non
\ee
\be
\int_X\m_\al\w\widetilde\m^\beta=\widehat{d}_\al\vphantom{\widehat{d}}^\beta,\qquad\int_X\om_a\w\widetilde\om^b=d_a\vphantom{d}^b.
\ee

In odd degree we will have both odd and even forms, and, since the volume form is even, we can construct a symplectic basis for each of $H^3_+(X)$ and $H^3_-(X)$.  For $H^3_+(X)$, we will have $a_K$, $b^K$, and for $H^3_-(X)$ we will have $\mathcal{A}_k$, $\mathcal{B}^k$.  The nonvanishing intersections are
\be
\int_Xa_K\w b^J=\d_K^J,\qquad\int_X\mathcal{A}_k\w\mathcal{B}^j=\d_k^j.
\ee

For the O3/O7 case, the index $K$ can take values $1\le K\le h^{2,1}_+$ and $k$ can run over $0\le k\le h^{2,1}_-$, with the extra index accounting for the fact that $H^{(3,0)}(X)\oplus H^{(0,3)}(X)$ is odd, while similarly for O5/O9 we have
$0\le K\le h^{2,1}_+$, $1\le k\le h^{2,1}_-$.

\subsection{The O3/O7 case}
\label{O3O7Case}

In this case the orientifold action requires that the holomorphic three form be odd and the K\"ahler form be even under $\s$.  Also, the $B$-field should be odd, as should the R-R fields $C_2$ and $C_6$, while the R-R fields $C_0$ and $C_4$ should be even.  With these projections we have the expansions
\be
\Om=\mathcal{Z}^k\mathcal{A}_k-\mathcal{F}_k\mathcal{B}^k,\qquad J=v^\al\m_\al,\qquad B=u^a\om_a,
\ee
\be
C_0,\qquad C_2=c^a\om_a,\qquad C_4=\rho_\al\widetilde\m^\al+A^K\w a_K,\qquad C_6=0.\non
\ee
Here we have not included any fluxes, nor any fields related to these by the self-duality of the R-R five-form field strength in IIB.  In fact it is easy to account for the latter; we would just add an extra piece to $C_4$,
\be
C_4'=\chi^\al\w\m_\al+\widetilde A_K\w b^K,
\ee
where $\chi^\al$ is a two-form potential in spacetime which is dual to the scalar field $\rho_\al$, and where $\widetilde A_K$ is the magnetic dual gauge field to $A^K$.

It turns out that the most convenient way to express these moduli is to follow~\cite{Benmachiche:2006df}\footnote{Note that our convention differs from~\cite{Benmachiche:2006df} in the sign of the $B$-field.} and define
\bea
\Phi_c^{ev} &=& e^B\w C_{RR}^{(0)}+ie^{-\phi}\Re\lp e^{B+iJ}\rp\non\\
&=& \lp C_0+ie^{-\phi}\rp+\lp C_2+\lp C_0+ie^{-\phi}\rp B\rp\\
&& +\lp C_4^{(0)}+C_2\w B+\hlf \lp C_0+ie^{-\phi}\rp B\w B-\frac{i}{2}e^{-\phi}J\w J\rp\non\\
&=& \tau+G^a\om_a+T_\al\widetilde\m^\al.
\eea
In this expression a superscript $(0)$ means that only the spacetime scalar part of an expansion is taken, and $C_{RR}=C_0+C_2+C_4$.  The expansion coefficients,
\bea
\tau &=& C_0+ie^{-\phi},\non\\
G^a &=& c^a+\tau u^a,\\
T_\al &=& \rho_\al+\lp\widehat{d}^{-1}\rp_\al\vphantom{\lp\widehat{d}^{-1}\rp}^\beta\lp-\frac{i}{2}e^{-\phi}\k_{\beta\g\d}v^\g v^\d+\widehat\k_{\beta\,ab}\lp c^au^b+\hlf\tau u^au^b\rp\rp,\non
\eea
turn out to be a nice basis for some of the complex scalar fields in four dimensions.  The remaining complex scalars are obtained from the fields $\mathcal{Z}^k$.  In fact the $\mathcal{Z}^k$ form a good projective basis, and we can use $z^k=\mathcal{Z}^k/\mathcal{Z}^0$, $1\le k\le h^{2,1}_-$, as a basis for the actual complex structure moduli.

The K\"ahler potential for these fields is then given by
\be
K=-\ln\ls i\int_X\Om\w\bar\Om\rs-4\ln\ls-i\lp\tau-\bar\tau\rp\rs-2\ln\ls 2\mathcal{V}_6\rs,
\ee
where the volume
\be
\mathcal{V}_6=\frac{1}{6}\int_XJ^3=\frac{1}{6}\k_{\al\beta\g}v^\al v^\beta v^\g
\ee
is implicitly viewed as a function of $T_\al$, $\tau$, and $G^a$.

The holomorphic gauge kinetic couplings can also be calculated~\cite{Grimm:2004uq}, though not as explicitly as in the IIA case.  The procedure is to consider the expansion of the holomorphic three-form before the orientifold projection
\be
\Om^{(0)}=\mathcal{Z}^k\mathcal{A}_k-\mathcal{F}_k\mathcal{B}^k+\mathcal{X}^Ka_K-\mathcal{G}_Kb^K,
\ee
where $\mathcal{F}_k$ and $\mathcal{G}_K$ are both considered to be functions of $\mathcal{Z}^k$ and $\mathcal{X}^K$.  Then the electric gauge kinetic couplings are given by
\be
f_{KJ}=-\frac{i}{2}\frac{\p}{\p\mathcal{X}^K}\mathcal{G}_J|_{\mathcal{X}^K=0}.
\ee
It can be shown~\cite{Grimm:2004uq} that $f_{KJ}$ are holomorphic functions of the complex structure moduli $z^k$.

The magnetic gauge kinetic couplings can also be computed by simply interchanging $a_K$ and $b^K$ (by a symplectic rotation) in the computation above.

Next we would like to include also a general set of fluxes.  In the R-R sector we can have
\be
F_3=m^k\mathcal{A}_k+e_k\mathcal{B}^k.
\ee

In the NS-NS sector we again introduce the fluxes $H_{ijk}$, $\om^i_{jk}$, $Q^{ij}_k$, and $R^{ijk}$ (with the same caveats as before regarding which fluxes can be obtained from known ten-dimensional constructions).  The Bianchi identities are still as given in (\ref{IIANongeometricBianchis}).  It is again convenient to define cohomological parameters, which in our new basis are
\be
H=p^k\mathcal{A}_k+p_k\mathcal{B}^k,\non
\ee
\be
\label{CohomologicalBasisFluxes}
\om\cdot\m_\al=\widehat{r}_\al^Ka_K+\widehat{r}_{\al K}b^K,\qquad\om\cdot\om_a=r_a^k\mathcal{A}_k+r_{ak}\mathcal{B}^k,
\ee
\be
Q\cdot\widetilde\m^\al=\widehat{q}^{\al k}\mathcal{A}_k+\widehat{q}^\al_k\mathcal{B}^k,\qquad Q\cdot\widetilde\om^a=q^{aK}a_K+q^a_Kb^K,\non
\ee
\be
R\cdot\varphi=s^Ka_K+s_Kb^K.\non
\ee
Note the abuse of notation here; fluxes with upper $H^3(X)$ indices (i.e. $K$ or $k$) are distinct from and independent of fluxes with lower $H^3(X)$ indices.  In other words we can turn on either, both, or neither of $p^k$ and $p_k$.  The discouraged reader should rest assured that this situation will not propagate throughout our entire analysis; shortly we will argue that all of the fluxes with upper $H^3(X)$ indices can consistently be set to zero.

We also have the nonvanishing actions
\be
H\w\mathcal{A}_k=-f^{-1}p_k\varphi,\qquad H\w\mathcal{B}^k=f^{-1}p^k\varphi,\non
\ee
\be
\om\cdot a_K=\lp\widehat{d}^{-1}\rp_\al\vphantom{\lp\widehat{d}^{-1}\rp}^\beta\widehat{r}_{\beta K}\widetilde\m^\al,\qquad\om\cdot b^K=-\lp\widehat{d}^{-1}\rp_\al\vphantom{\lp\widehat{d}^{-1}\rp}^\beta\widehat{r}_\beta^K\widetilde\m^\al,\non
\ee
\be
\om\cdot\mathcal{A}_k=\lp d^{-1}\rp_a\vphantom{\lp d^{-1}\rp}^br_{bk}\widetilde\om^a,\qquad\om\cdot\mathcal{B}^k=-\lp d^{-1}\rp_a\vphantom{\lp d^{-1}\rp}^br_b^k\widetilde\om^a,
\ee
\be
Q\cdot a_K=-\lp d^{-1}\rp_a\vphantom{\lp d^{-1}\rp}^bq^a_K\om_b,\qquad Q\cdot b^K=\lp d^{-1}\rp_a\vphantom{\lp d^{-1}\rp}^bq^{aK}\om_b,\non
\ee
\be
Q\cdot\mathcal{A}_k=-\lp\widehat{d}^{-1}\rp_\al\vphantom{\lp\widehat{d}^{-1}\rp}^\beta\widehat{q}^\al_k\m_\beta,\qquad Q\cdot\mathcal{B}^k=\lp\widehat{d}^{-1}\rp_\al\vphantom{\lp\widehat{d}^{-1}\rp}^\beta\widehat{q}^{\al k}\m_\beta,\non
\ee
\be
R\cdot a_K=f^{-1}s_K1,\qquad R\cdot b^K=-f^{-1}s^K1.\non
\ee

Once again, we can define an operator $\mathcal{D}$, as in (\ref{IIAD}), by using the same component-wise action of $H\w$ and the actions of the remaining fluxes from equations (\ref{omcdotA}), (\ref{QAction}), and (\ref{RAction}).  The Bianchi identities can still be derived by enforcing $\mathcal{D}^2=0$, and by demanding that $\mathcal{D}^2$ vanish on our cohomological basis we get a subset of the Bianchi identities, but naturally expressed in terms of the parameters defined above as
\bea
p^k\widehat{q}^\al_k-p_k\widehat{q}^{\al k}=\widehat{r}_\al^Ks_K-\widehat{r}_{\al K}s^K &=& 0,\qquad\forall\al,\non\\
p^kr_{ak}-p_kr_a^k=q^{aK}s_K-q^a_Ks^K &=& 0,\qquad\forall a,\non\\
\label{IIBMutualLocalityBianchis}
\widehat{r}_{[\al}^K\widehat{r}_{\beta] K}=\widehat{q}^{[\al}_k\widehat{q}^{\beta]k} &=& 0,\qquad\forall\al,\beta,\\
\widehat{r}_\al^Kq_{bK}-\widehat{r}_{\al K}q_b^K=\widehat{q}^{\al k}r_{bk}-\widehat{q}^\al_kr_b^k &=& 0,\qquad\forall\al,b,\non\\
r_{[a}^kr_{b]k}=q^{[a}_Kq^{b]K} &=& 0,\qquad\forall a,b,\non
\eea
and
\be
\label{IIBkJCohomologicalBianchi}
f^{-1}p_ks_J-\lp d^{-1}\rp_a\vphantom{\lp d^{-1}\rp}^br_{bk}q^a_J+\lp\widehat{d}^{-1}\rp_\al\vphantom{\lp\widehat{d}^{-1}\rp}^\beta\widehat{q}^\al_k\widehat{r}_{\beta J}=0,\qquad\forall k,J,
\ee
and where (\ref{IIBkJCohomologicalBianchi}) also holds with either or both of the indices $k$ and $J$ raised.

The equations (\ref{IIBMutualLocalityBianchis}) have a very useful interpretation.  They tell us that the vectors $(\widehat{r}_\al^K,\widehat{r}_{\al K})$, $(q^{aK},q^a_K)$, and $(s^K,s_K)$, are a symplectically orthogonal set with respect to the symplectic basis $(a_K,b^K)$, and that $(p_k,p^k)$, $(r_{ak},r_a^k)$, and $(\widehat{q}^\al_k,\widehat{q}^{\al k})$, are a symplectically orthogonal set with respect to $(\mathcal{A}_k,\mathcal{B}^k)$.  But given any collection of symplectically orthogonal vectors  there exists a symplectic transformation which rotates them so that they all lie within a canonical Lagrangian subspace.  In other words, we can rotate our symplectic basis so that all vector components with an upper index vanish, and we are left with only the components carrying a lower index.  This procedure is used, for example, in the case of dyonic charge vectors.  The symplectic orthogonality conditions are then called mutual locality of the different charged fields, and when they are satisfied we may rotate our electric and magnetic gauge fields so that all charges are purely electric.  Thus we are free to assume that all of our fluxes have only lower $H^3(X)$ indices and that all components with upper indices vanish.

Note that we are glossing over an important point, namely that if we want to map our integral symplectic basis into another integral basis, then our rotation should sit inside of $\Sp(n;\Z)$.  In this case our procedure is only possible if for each charge vector $q_A^{(i)}=(q_k^{(i)};q^{k\,(i)})$, the ratios of all components are rational, i.e. if there exists some real number $g^{(i)}\ge 0$ and integers $n_A^{(i)}$ such that $q_A^{(i)}=g^{(i)}n_A^{(i)}$ (note that we do not require any relations here between different charge vectors).  For instance, for a single such vector, there exists a rotation in $\Sp(n;\Z)$ sending $q_A$ to $q_A'=(g_{\mathrm{max}},0,\ldots,0;0,\ldots,0)$, where $g_{\mathrm{max}}$ is the largest real number $g$ with the above property (if all the original $q_A$ were integers, then $g_{\mathrm{max}}=\operatorname{gcd}(q_A)$).  If some vector does not have this property, there is no $\Sp(n;\Z)$ rotation to do what we need.  We can still, however, use a $\Sp(n)$ rotation to eliminate the unwanted fluxes, derive any formulae in the simplified situation, and then rotate back to the integral symplectic basis.

For some of the fluxes these issues are merely technicalities (because, as mentioned, we can always undo our rotation at the end), but below we will argue that some of these vectors are in fact physical charges (under the electric and dual magnetic fields of the four-dimensional effective theory) and thus should, for quantum consistency, be quantized, at least when everything has been correctly normalized.  Unfortunately, to settle this question one needs to understand the correct quantization condition for these generalized NS-NS fluxes.  In~\cite{Ihl:2007ah} it is shown how to do this in a broad class of examples, and the resulting quantization conditions are found to be quite nontrivial.  A similar construction can easily be done for IIB toroidal orientifolds, and the same conclusions will hold, but a proof of charge quantization in even this class of examples eludes us, though it holds in all cases that we have checked.  In the general situation, outside of this class, it is not clear to us how to even check the result.

Neglecting these issues, this simplification means that equations (\ref{IIBMutualLocalityBianchis}) are automatically satisfied, and (\ref{IIBkJCohomologicalBianchi}) reduces to just one set of equations, rather than four.

Let us turn now to potential tadpoles for space-filling R-R form fields.  By T-dualizing the type IIA tadpole constraint (\ref{IIAGeneralTadpole}) we arrive at the IIB tadpole constraint in the presence of general fluxes,
\be
\label{IIBTadpole}
\mathcal{D} F_3|_{(9-p)-form}+\ls\d_{\mathrm{D_p-branes}}\rs=2^{p-5}\ls\d_{\mathrm{O_p-planes}}\rs.
\ee
In section~\ref{MainExample}, we will briefly discuss how one computes the O-plane contributions above.  For now, let us consider this equation degree by degree.

First we have the $C_4$ tadpole.  Since we are in the O3/O7 class of orientifolds, there is certainly the possibility of an orientifold group element with a real codimension six fixed locus, i.e. an O3-plane.  Additionally, we can have spacetime-filling D3-branes sitting at points on the internal manifold.  With a change of orientation, we can also have anti-D3-branes, but these will break supersymmetry. In total, the constraint reads
\be
H\w F_3+\ls\d_{D3}\rs=\frac{1}{4}\ls\d_{O3}\rs,
\ee
or in components (integrating over $X$),
\be
-p_km^k+N_{D3}=\frac{1}{4} N_{O3}.
\ee

Next we can consider the potential $C_6$ tadpole.  Since $C_6$ needs to be odd under the orientifold projection, we can only get contributions proportional to odd four-forms, i.e. the $\widetilde\om^a$.  There can be no contribution to this tadpole from O-planes, since O5-planes are not consistent with the O3/O7 class of orientifolds.  In principle we can have D5-branes contributing, but they will necessarily break supersymmetry.  To see this, recall that a supersymmetric two-cycle in our compactification manifold should be one that is calibrated by the K\"ahler form $J$.  But since the orientifold projection picks out an odd two-cycle and forces $J$ to be an even two-form, $J$ clearly vanishes when pulled back to the D5 worldvolume (equivalently, $J\w\widetilde\om^a=0$).  Our condition is hence,
\be
-\om \cdot F_3+\ls\d_{D5}\rs=0,
\ee
where any localized contribution breaks SUSY.  Thus, in a supersymmetric vacuum, we have, in components,
\be
r_{ak}m^k=0.
\ee

We move on to $C_8$, and find the result
\be
Q\cdot F_3+\ls\d_{D7}\rs=4\ls\d_{O7}\rs,
\ee
or
\be
-\widehat{q}^\al_km^k+N_{D7}^\al=4 N_{O7}^\al.
\ee

And finally, the $C_{10}$ tadpole is absent, since it must be odd under the orientifold projection, but there is no odd six-cycle on the internal manifold, or equivalently, no odd zero-form.

Let us remark briefly on a special case of the above constraints.  If there are no localized sources (we will present such an example in section \ref{Examples}) or if the localized sources are engineered to cancel amongst themselves (i.e. any O-plane charge is cancelled by adding D-branes), then the tadpole constraints above make the simple statement that the R-R charge vector $(e_k,m^k)$ is again symplectically orthogonal to our various NS-NS vectors, which after our earlier rotations are simply $(p_k,0)$, $(r_{ak},0)$, and $(\widehat{q}^\al_k,0)$.  In this case there will again be a symplectic rotation which will eliminate the $m^k$ components of $F_3$, leaving only the $e_k$.  The quantization issues discussed above will still be present, but we will not repeat the details.  If the flux contribution to the tadpoles does not vanish however, but rather is required to cancel local source contributions, then this argument does not apply.

Next, we turn to the superpotential.  Known results from solutions with $H$-flux and torsion allow us to use T-duality to write down the superpotential with general NS-NS fluxes.  We find~\cite{Benmachiche:2006df}
\be
\label{GeneralIIBSuperpotential}
W=\int_X\lp F_3+\mathcal{D}\Phi^{ev}_c\rp\w\Om.
\ee
Computing explicitly, we find that
\be
\mathcal{D}\Phi^{ev}_c=\lp p_k\tau+r_{ak}G^a+\widehat{q}^\al_kT_\al\rp\mathcal{B}^k,
\ee
so doing the integration, we find our superpotential to be
\be
W=-m^k\mathcal{F}_k-\ls e_k+p_k\tau+r_{ak}G^a+\widehat{q}^\al_kT_\al\rs\mathcal{Z}^k.
\ee
Note particularly that $W$ is linear in the moduli $\tau$, $G^a$, and $T_\al$.  Also, observe that in the presence of the nongeometric $\widehat{q}$ fluxes, the superpotential does depend on the volume moduli of the compactificatoin, meaning that there is at least a chance to stabilize everything at tree level.

Finally, we turn to the D-terms.  Proceeding as in the IIA case, we note that gauge transformations of the electric gauge fields $A^K$ and the magnetic gauge fields $\widetilde{A}_K$ are generated by
\bea
C_{RR} &\longrightarrow& C_{RR}+\mathcal{D}\lp\lambda^Ka_K+\widetilde{\lambda}_Kb^K\rp\non\\
&& = \lp C_0-f^{-1}s_K\lambda^K\rp+\lp c^a-\lp d^{-1}\rp_b\vphantom{\lp d^{-1}\rp}^aq^b_K\lambda^K\rp\om_a+\lp\rho_\al-\lp\widehat{d}^{-1}\rp_\al\vphantom{\lp\widehat{d}^{-1}\rp}^\beta\widehat{r}_{\beta K}\lambda^K\rp\widetilde{\m}^\al\non\\
&& \qquad+\lp A^K+d\lambda^K\rp\w a_K+\lp\widetilde{A}_K+d\widetilde{\lambda}_K\rp\w b^K.
\eea

Thus the fields $\tau$, $G^a$, and $T_\al$ can all potentially get variations under electric gauge transformations by turning on our general fluxes.  Observe that if we hadn't performed a symplectic rotation of the general fluxes, then both electric and magnetic charges would have been possible, and that indeed the symplectic vectors discussed above would be precisely the dyonic charge vectors, as promised.  Note also that the fluxes which contribute to charges of these fields are a complementary set to the fluxes which can appear in the superpotential and tadpole constraints.

The D-terms which result from these variations are
\bea
D_K &=& -i\ls f^{-1}s_K\p_\tau K+\lp d^{-1}\rp_b\vphantom{\lp d^{-1}\rp}^aq^b_K\p_aK+\lp\widehat{d}^{-1}\rp_\al\vphantom{\lp\widehat{d}^{-1}\rp}^\beta\widehat{r}_{\beta K}\p^\al K\rs\\
&=& \frac{e^\phi}{2 \mathcal{V}_6} \ls\lp \mathcal{V}_6-\hlf\lp\widehat{\k}vu^2\rp\rp f^{-1}s_K+\lp d^{-1}\rp_a\vphantom{\lp d^{-1}\rp}^b\widehat{\k}_{\al bc}v^\al u^cq^a_K-v^\al\widehat{r}_{\al K}\rs.\non
\eea

We will see how this works in a specific example below.

\subsection{The O5/O9 case}
\label{O5O9Case}

This case is quite similar to the previous case, so we shall be fairly brief in our description.  The holomorphic involution $\s$ now satisfies $\s^*\Om=\Om$, and the projection on the R-R sector is reversed relative to the O3/O7 case (because the projection is no longer accompanied by a factor of $(-1)^{F_L}$), so we are left with the expansions
\be
\Om=\mathcal{Z}^Ka_K-\mathcal{F}_Kb^K,\qquad J=v^\al\m_\al,\qquad B=u^a\om_a,
\ee
\be
C_0=0,\qquad C_2=c^\al\m_\al,\qquad C_4=\rho_a\widetilde{\om}^a+A^k\w\mathcal{A}_k,\qquad C_6=\g\varphi.\non
\ee
There are of course also the dual pieces of $C_4$.  Also, the field $\g$ in $C_6$ is dual to a spacetime two-form field from $C_2$, but we prefer to work with spacetime scalars in our description.

It is again convenient to introduce the formal sum of forms~\cite{Benmachiche:2006df}
\bea
\Phi_c^{ev} &=& e^B\w C_{RR}^{(0)}+ie^{-\phi}\Im\lp e^{B+iJ}\rp\non\\
&=& \lp C_2+ie^{-\phi}J\rp+\lp C_4^{(0)}+C_2\w B+ie^{-\phi}B\w J\rp\\
&& +\lp C_6+C_4^{(0)}\w B+\hlf C_2\w B\w B+ie^{-\phi}\lp-\frac{1}{6}J\w J\w J+\hlf J\w B\w B\rp\rp\non\\
&=& t^\al\m_\al+L_a\widetilde{\om}^a+S\varphi,
\eea
with
\bea
t^\al &=& c^\al+ie^{-\phi}v^\al,\non\\
L_a &=& \rho_a+\lp d^{-1}\rp_a\vphantom{\lp d^{-1}\rp}^b\widehat{\k}_{\al\,bc}t^\al u^c,\\
S &=& \g+f^{-1}\ls d_a\vphantom{d}^b\rho_bu^a+\hlf\widehat{\k}_{\al\,ab}t^\al u^au^b-\frac{i}{6}e^{-\phi}\k_{\al\beta\g}v^\al v^\beta v^\g\rs.\non
\eea
These fields, $t^\al$, $L_a$, and $S$, are good holomorphic coordinates on the moduli space, and should be combined with projective coordinates for the complex structure deformations,
\be
z^K=\mathcal{Z}^K/\mathcal{Z}^0,\qquad 1\le K\le h^{2,1}_K.
\ee

The K\"ahler potential is given by the same expression as before,
\be
K=-\ln\ls i\int_X\Om\w\bar\Om\rs-4\ln\ls e^{-\phi}\rs-2\ln\ls 8\mathcal{V}_6\rs,
\ee
but should now be viewed as an implicit function of $z^K$, $t^\al$, $L_a$, and $S$.

To compute the gauge kinetic couplings we follow the same procedure as before, constructing the three-form before the orientifold projection
\be
\Om^{(0)}=\mathcal{Z}^Ka_K-\mathcal{F}_Kb^K+\mathcal{X}^k\mathcal{A}_k-\mathcal{G}_k\mathcal{B}^k,
\ee
with $\mathcal{F}_K$ and $\mathcal{G}_k$ considered as functions of $\mathcal{Z}^K$ and $\mathcal{X}^k$.  We then have
\be
f_{kj}=-\frac{i}{2}\frac{\p}{\p\mathcal{X}^k}\mathcal{G}_j|_{\mathcal{X}^k=0}.
\ee

Next we turn to fluxes.  In the R-R sector, we have simply
\be
F_3=m^Ka_K+e_Kb^K.
\ee

In the NS-NS sector we have precisely the same expansion (\ref{CohomologicalBasisFluxes}) as before, with the same Bianchi identities (\ref{IIBMutualLocalityBianchis}) and (\ref{IIBkJCohomologicalBianchi}), and where again we can rotate our symplectic basis so that only ``electric'' fluxes remain.

We again expect the tadpole to be given by (\ref{IIBTadpole}), but now the degree-by-degree comparison will be different.

There is no possible $C_4$ tadpole, since a space-filling $C_4$ field is projected out by the orientifold.

There are possible $C_6$ tadpoles,
\be
-\om\cdot F_3+\ls\d_{D5}\rs=\ls\d_{O5}\rs,
\ee
or
\be
-\widehat{r}_{\al K}m^K+N^{D5}_\al=N^{O5}_\al.
\ee

For $C_8$ we have
\be
Q\cdot F_3+\ls\d_{D7}\rs=0,
\ee
with the caveat that any D7-branes surviving the orientifold projection are necessarily non-supersymmetric.  In components, in the supersymmetric case, we find
\be
q^a_Km^K=0.
\ee

Finally, there is a potential $C_{10}$ tadpole
\be
-R\cdot F_3+\ls\d_{D9}\rs=16 \ls\d_{O9}\rs,
\ee
or
\be
-s_Km^K+N_{D9}=16 N_{O9}.
\ee

Once again, in the absence of localized sources, a further symplectic rotation can also eliminate the $m^K$ components of $F^3$.

The superpotential is actually given by the same general expression (\ref{GeneralIIBSuperpotential}), but where the expansion now reads
\be
W=-m^K\mathcal{F}_K-\ls e_K+t^\al\widehat{r}_{\al K}+q^a_KL_a+s_KS\rs\mathcal{Z}^K.
\ee

It remains only to compute the D-terms.  We find that under the standard gauge variation,
\bea
C_{RR} &\longrightarrow& C_{RR}+\mathcal{D}\lp\lambda^k\mathcal{A}_k+\widetilde{\lambda}_k\mathcal{B}^k\rp\non\\
&& =\lp c^\al-\lp\widehat{d}^{-1}\rp_\beta\vphantom{\lp\widehat{d}^{-1}\rp}^\al\widehat{q}^\beta_k\lambda^k\rp\m_\al+\lp\rho_a-\lp d^{-1}\rp_a\vphantom{\lp d^{-1}\rp}^br_{bk}\lambda^k\rp\widetilde{\om}^a+\lp\g-f^{-1}p_k\lambda^k\rp\varphi\non\\
&& \quad+\lp A^k+d\lambda^k\rp\w\mathcal{A}_k+\lp\widetilde{A}_k+d\widetilde{\lambda}_k\rp\w\mathcal{B}^k.
\eea
Finally, we can compute the D-terms,
\bea
D_k &=& -i\ls\lp\widehat{d}^{-1}\rp_\beta\vphantom{\lp\widehat{d}^{-1}\rp}^\al\widehat{q}^\beta_k\p_\al K+\lp d^{-1}\rp_a\vphantom{\lp d^{-1}\rp}^br_{bk}\p^aK+f^{-1}p_k\p_SK\rs\non\\
&=& \frac{e^\phi}{2\mathcal{V}_6}\left\{\hlf\lp\widehat{d}^{-1}\rp_\beta\vphantom{\lp\widehat{d}^{-1}\rp}^\al\widehat{q}^\beta_k\lp\k_{\al\g\d}v^\g v^\d-\widehat{\k}_{\al\,ab}u^au^b\rp+r_{ak}u^a-p_k\right\}.
\eea

Note that the last term above, proportional to $p_k$, matches the result found in~\cite{Grimm:2004uq}.

\section{Examples}
\label{Examples}

In this section we will work out explicitly the example of D-terms arising in type IIB supergravity compactified on the orbifold $T^6/\Z_4$ with an $O3/O7$ orientifold. A similar example of an $O5/O9$ orientifold can be obtain by slightly modifying the holomorphic involution as explained below. For a completely worked out example in type IIA see~\cite{Ihl:2007ah}.

Before we launch into a description of the example we have in mind, it is worth briefly commenting about why IIB examples that exhibit D-terms are somewhat difficult to find.  Consider the O3/O7 case.  In order to have a possibility for D-terms we need to have $h^{2,1}_+>0$, so that we have four-dimensional vectors, and we also need either nongeometric $s$- or $q$-fluxes, or else we need metric $\widehat{r}$-fluxes to act as charges.  In fact, since most studies of generalized NS-NS fluxes (present work included) have really focused on the untwisted sectors of a toroidal orientifold\footnote{For a way of including part of the untwisted sector see \cite{Cvetic:2007ju}.}, we actually want $h^{2,1}_{+\,\mathrm{untwisted}}>0$.  But nearly all standard examples of O3/O7 toroidal orientifolds use $\s=\mathcal{I}_6$, a reflection of all internal coordinates.  Under such an involution, of course all untwisted three-forms are odd.  Other common examples start with a factorized orbifold of $(T^2)^3$, and take an involution which reflects one of the two-tori, but here too one can show (assuming the orbifold didn't enjoy enhanced supersymmetry) that all untwisted three-forms are odd.  So we need to look for a slightly more involved example, which we will describe below.

\subsection{O3/O7 on $T^6/\Z_4$}
\label{MainExample}

We start by explicitly spelling out the orbifold and orientifold action and the resulting cohomology. Then we discuss the $H$, metric and non-geometric fluxes and how they map to cohomological parameters. Finally we write down explicitly the D-terms.

Let $z^1=x^1+i x^2+e^{\pi \ic/4} (x^3+\ic x^4)$, $z^2=x^3+i x^4+ e^{3 \pi \ic/4} (x^1+\ic x^2)$, and $z^3=x^5+\ic x^6$ be complex coordinates on the tori with the identifications $x^i = x^i +1$. The orientifold group is generated by a $\Z_4$ rotation
\be
\Theta:\lp z^1,z^2,z^3\rp\longrightarrow\lp \ic z^1,\ic z^2,-z^3\rp,
\ee
and the orbifold action is $\Om_p(-1)^{F_L}\s$, where the holomorphic involution $\s$ acts as
\be
\label{ExampleInvolution}
\s:\lp z^1,z^2,z^3\rp\longrightarrow\lp -e^{\pi \ic/4} z^1, e^{\pi \ic/4} z^2,-\ic z^3\rp.
\ee
Note that $\s^2=\Theta$, so the full orientifold group is in fact $\Z_8$.  More specifically, for those familiar with classifications of crystallographic actions on $T^6$, if we pair $\s$ with a reflection in all six coordinates, the element $\s\mathcal{I}_6$ generates the crystallographic group $\Z_{8-I}$ (see for instance the review~\cite{Reffert:2007im}).  This particular orientifold is discussed by~\cite{Rabadan:2000ma}.

We can now write down the untwisted cohomology of $T^6/\Z_4$, dividing further into subspaces which are even or odd under the involution $\s$. We start with the even cohomology, implicitly equating classes with their harmonic form representatives. There is one even zero form, namely the unit function $1$.  For two-forms, there are five independent $(1,1)$-forms invariant under the rotations: three even forms,
\bea
\m_1 &=& \frac{i}{4}\lp dz^1\w d\bar z^1 + dz^2 \w d\bar z^2 \rp= dx^1\w dx^2+dx^3 \w dx^4,\non\\
\m_2 &=& \frac{i}{2 \sqrt{2}}\lp dz^1\w d\bar z^1 - dz^2 \w d\bar z^2 \rp= dx^1\w dx^3 + dx^1 \w dx^4 -dx^2 \w dx^3 +dx^2 \w dx^4,\non\\
\m_3 &=& \frac{i}{2}dz^3\w d\bar z^3=dx^5\w dx^6,\non
\eea
and two odd forms
\bea
\om_1 &=& \frac{1-\ic}{4}\lp dz^1\w d\bar z^2+ \ic d\bar z^1 \w dz^2\rp = dx^1\w dx^3- dx^1\w dx^4 +dx^2 \w dx^3+dx^2 \w dx^4,\non\\
\om_2 &=& -\frac{e^{-\pi \ic/4}}{4}\lp dz^1\w d\bar z^2 - \ic d\bar z^1\w d z^2\rp = dx^1\w dx^2 - dx^3\w dx^4.\non
\eea
Similarly, for four-forms we have three even $(2,2)$-forms
\bea
\widetilde\m^1 &=& \m_1 \w\m_3,\qquad\widetilde\m^2=\m_2\w\m_3,\non \\
\widetilde\m^3 &=& dx^1 \w dx^2 \w dx^3 \w dx^4 = \frac{1}{2} \, \m_1\w\m_1=-\frac{1}{4} \, \m_2 \w \m_2= -\frac{1}{4} \, \om_1 \w \om_1= -\frac{1}{2} \, \om_2 \w \om_2,\non
\eea
and two odd $(2,2)$-form,
\be
\widetilde\om^1=\om_1 \w \m_3\qquad\widetilde\om^2=\om_2 \w \m_3.
\ee
Finally there is one six-form, which is even under the involution,
\be
\varphi=dx^1\w dx^2\w dx^3\w dx^4\w dx^5\w dx^6.
\ee
For the intersection numbers we find $f=\frac{1}{4}$, $\widehat d_\al\vphantom{\widehat{d}}^\beta =diag\lp \frac{1}{2},-1,\frac{1}{4}\rp$ and $d_a\vphantom{d}^b=diag\lp-1,-\frac{1}{2}\rp$. The only non vanishing components of the totally symmetric triple intersections are $\k_{113}=$~$\frac{1}{2},$ $\k_{223}=-1$ and $\widehat \k_{311}=-1, \, \widehat \k_{322}=-\frac{1}{2}$.

In particular, the K\"ahler form will be given by $J=v^1\m_1+v^2\m_2+v^3\m_3$, and the corresponding metric (in the absence of fluxes) is
\bea
ds^2 &=& v^1\lp (dx^1)^2+(dx^2)^2+(dx^3)^2+(dx^4)^2\rp+2v^2\lp dx^1dx^3-dx^1dx^4+dx^2dx^3+dx^2dx^4\rp\non\\
&& +v^3\lp (dx^5)^2+(dx^6)^2\rp.
\eea
The conditions that the metric be Euclidean signature are that $v^1>0$, $v^3>0$, and that $(v^1)^2>2(v^2)^2$.  The volume is
\be
\mathcal{V}_6=\frac{1}{4}v^3\lp\lp v^1\rp^2-2\lp v^2\rp^2\rp.
\ee

Next we have the odd cohomology.  It turns out that $H^1(X)$ and $H^5(X)$ are empty, so we need only describe the three-forms.  Since there are only four we drop the index and simply write
\bea
a &=& -\frac{\ic}{2} \lp dz^1 \w dz^2 \w d\bar z^3 - d\bar z^1 \w d\bar z^2 \w d z^3 \rp = -\chi^{136}+\chi^{145}+\chi^{235}+\chi^{246},\\\non
b &=& \frac{1}{2} \lp dz^1 \w dz^2 \w d\bar z^3 + d\bar z^1 \w d\bar z^2 \w d z^3 \rp= \chi^{135}-\chi^{245}+\chi^{146}+\chi^{236},\non\\
\mathcal{A} &=& \frac{1}{2} \lp dz^1 \w dz^2 \w dz^3 + d\bar z^1 \w d\bar z^2 \w d\bar z^3 \rp = \chi^{135}-\chi^{245}-\chi^{146}-\chi^{236},\non\\
\mathcal{B} &=& -\frac{\ic}{2} \lp dz^1 \w dz^2 \w dz^3 - d\bar z^1 \w d\bar z^2 \w d\bar z^3 \rp = \chi^{136}+\chi^{145}+\chi^{235}-\chi^{246}.
\eea
Here we use notation where $\chi^{145}=dx^1\w dx^4\w dx^5$, etc. The holomorphic three form
\be
\Omega = \frac{1}{\sqrt{2}} \, dz^1 \w dz^2 \w dz^3 = \mathcal{Z} \mathcal{A} - \mathcal{F}  \mathcal{B} = \frac{1}{\sqrt{2}} (\mathcal{A} + \ic \mathcal{B})
\ee
is odd under $\s$ so that we have the O3/O7 case.  The normalization has been chosen so that $i\int_X\Om\w\bar\Om=1$, and the phase chosen so that $\mathcal{Z}^0$ is real and positive, but these are arbitrary choices.  Note that there are no complex structure moduli in this example.

We will now enumerate the general (untwisted) NS-NS fluxes that are consistent with the orientifold action.  First we expand $H = p^1 \mathcal{A} + p_1 \mathcal{B}$ where the parameters are $p^1 = H_{135}=- H_{245}=- H_{146}= - H_{236}$ and $p_1 = H_{136}= H_{145}= H_{235}= - H_{246}$.

Now proceed analogously for the other fluxes arising in the NS-NS sector. Imposing invariance under the orientifold group, we find that we are left with ten independent metric fluxes,
\bea
& \omega^1_{15} &=- \omega^2_{25} = -\omega^3_{36} =\omega^4_{46},\non\\
& \omega^1_{16} &=- \omega^2_{26}=\omega^3_{35} =-\omega^4_{45},\non\\
& \omega^1_{25} &= \omega^2_{15} =- \omega^3_{46} =-\omega^4_{36},\non\\
& \omega^1_{26} &=\omega^2_{16} =\omega^3_{45} =\omega^4_{35},\non\\
& \omega^1_{35} &=- \omega^2_{45} =- \omega^3_{26} =-\omega^4_{16},\non\\
& \omega^1_{36} &=- \omega^2_{46} =\omega^3_{25} =\omega^4_{15},\non\\
& \omega^1_{45} &=\omega^2_{35} =\omega^3_{16} =- \omega^4_{26},\non\\
& \omega^1_{46} &= \omega^2_{36} =- \omega^3_{15} =\omega^4_{25},\non\\
& \omega^5_{13} &=- \omega^5_{24} = \omega^6_{14} =\omega^6_{23},\non\\
& \omega^5_{14} &= \omega^5_{23} =- \omega^6_{13} =\omega^6_{24},\non
\eea
where we can use the ten fluxes in the left-hand column as representatives.\\
In terms of $r$-matrices, we find
\be
r_{a}^1 =\lp \begin{matrix} -\om^1_{15} -\om^1_{16} - \om^1_{25} + \om^1_{26} \\ - \om^1_{36} - \om^1_{45} \end{matrix}\rp, \qquad r_{a1} =\lp \begin{matrix} \om^1_{15} -\om^1_{16} - \om^1_{25} -\om^1_{26} \\ \om^1_{35} -\om^1_{46} \end{matrix}\rp,
\ee
\be
\widehat r_{\alpha}^1 = \lp\begin{matrix} \om^1_{35} + \om^1_{46} \\ -\om^1_{15} + \om^1_{16} - \om^1_{25} -\om^1_{26} \\ -\om^5_{13}\end{matrix}\rp, \qquad \widehat r_{\alpha 1}=\lp\begin{matrix} \om^1_{36} - \om^1_{45}\\ -\om^1_{15} -\om^1_{16} +\om^1_{25} -\om^1_{26}\\  \om^5_{14}\end{matrix}\rp.
\ee
Note that there is a one-to-one correspondence between the independent fluxes $\om^i_{jk}$ and the entries of $r$ and $\widehat r$. If we consider only these metric fluxes and set $r_{a}^1=r_{a1}=0$ we are left with the following Bianchi identities
\be\label{eq:metricBianchi}
\widehat r_{\gamma 1} \, \widehat r_{3 1} + \widehat r_{\gamma}^1 \, \widehat r_{3}^1 =0, \, \gamma =1,2, \quad (\widehat r_{11})^2 +(\widehat r_{1}^1)^2 - \frac{(\widehat r_{21})^2}{2}-\frac{(\widehat r_{2}^1)^2}{2}=0, \quad \widehat r_{[\alpha}^1 \,  \widehat r_{\beta]1} =0.
\ee
Note that only the last Bianchi identity arises from demanding that $\mathcal{D}^2$ vanishes when acting on the invariant forms given above (cf. \eqref{IIBMutualLocalityBianchis}). One solution to \eqref{eq:metricBianchi} which gives a D-term is for example to turn on only the components $\widehat r_3^1$ and $\widehat r_{31}$.

The $Q$-fluxes which survive the orientifold projection are
\bea
&Q^{13}_5&=-Q^{14}_6=-Q^{23}_6=-Q^{24}_5,\non\\
&Q^{13}_6&=Q^{14}_5=Q^{23}_5=-Q^{24}_6,\non\\
&Q^{15}_1&=-Q^{25}_2=Q^{36}_3=-Q^{46}_4,\non\\
&Q^{15}_2&=Q^{25}_1=Q^{36}_4=Q^{46}_3,\non\\
&Q^{15}_3&=-Q^{25}_4=Q^{36}_2=Q^{46}_1,\non\\
&Q^{15}_4&=Q^{25}_3=-Q^{36}_1=Q^{46}_2,\non\\
&Q^{16}_1&=-Q^{26}_2=-Q^{35}_3=Q^{45}_4,\non\\
&Q^{16}_2&=Q^{26}_1=-Q^{35}_4=-Q^{45}_3,\non\\
&Q^{16}_3&=-Q^{26}_4=-Q^{35}_2=-Q^{45}_1,\non\\
&Q^{16}_4&=Q^{26}_3=Q^{35}_1=-Q^{45}_2,\non
\eea
where we take the ten fluxes in the left-hand column as representatives.
In terms of $q$-matrices, we find
\be
q^{a1}=\lp\begin{matrix} -Q^{15}_1 - Q^{15}_2 +Q^{16}_1 - Q^{16}_2 \\ -Q^{15}_4 +Q^{16}_3 \end{matrix}\rp, \qquad q^a_1=\lp\begin{matrix} Q^{15}_1 - Q^{15}_2 +Q^{16}_1 + Q^{16}_2\\ Q^{15}_3 + Q^{16}_4 \end{matrix}\rp,
\ee
\be
\widehat q^{\alpha1} =\lp\begin{matrix} -Q^{15}_3+ Q^{16}_4 \\ Q^{15}_1 + Q^{15}_2 +Q^{16}_1 - Q^{16}_2\\ Q^{13}_5 \end{matrix}\rp, \qquad \widehat q^\alpha_1=\lp\begin{matrix} -Q^{15}_4- Q^{16}_3  \\ -Q^{15}_1 + Q^{15}_2 +Q^{16}_1 + Q^{16}_2 \\Q^{13}_6 \end{matrix}\rp.
\ee
Note that there is a one-to-one correspondence between the independent fluxes $Q^i_{jk}$ and the entries of $q$ and $\widehat q$.

Finally we find $s^1 = -R^{135} = R^{245}= -R^{146}= -R^{236}$ and $s_1 = -R^{136} = R^{145}= R^{235}= R^{246}$.

For a way to understand from a 10-dimensional point which of the NS-NS fluxes discussed so far can be turned on see~\cite{Ihl:2007ah}.  The base-fiber constructions described there can easily be adapted to the IIB case to give a large class of ten dimensional constructions with $H$-flux, metric flux, and $Q$-flux.

If we demand that $\mathcal{D}^2$ vanishes we find the Bianchi identities derived above in equation \eqref{IIANongeometricBianchis}. This simplifies if we do a symplectic rotation so that only electric fluxes with lower $k, K$ indices are non-zero. Then we have to satisfy
\begin{align}
&p_1 q^c_1 =0, \, p_1 \widehat q^\gamma_1 =0, \, r_{c1} \widehat r_{31} =0, \, \widehat r_{\gamma 1} \widehat r_{31} =0, \, c, \gamma = 1,2, \\
&4 p_1 \widehat q^3_1 +\frac{(r_{11})^2}{2} +(r_{21})^2 + (\widehat r_{11})^2 - \frac{(\widehat r_{21})^2}{2}=0,\\
&s_1 \widehat r_{\alpha1} =0, \, \alpha=1,2,3, \, \widehat q^\gamma_1 \widehat q^3_1 =0, s_1 r_{c1} - q^c_1 \widehat q^3_1 =0, \,  \, c, \gamma = 1,2, \\
& \frac{(q^1_1)^2}{2} +(q^2_1)^2 + (\widehat q^1_1)^2 - \frac{(\widehat q^2_1)^2}{2}=0,\\
& r_{c1} \widehat q^\gamma_1 =0, \, \widehat r_{\gamma 1} q^c_1 =0, q^c_1 \widehat r_{31}=0 \, c,\gamma =1,2, \, p_1 s_1 =0, \, \widehat r_{31} \widehat q^3_1 =0,\\
&r_{11} q^1_1 + 2 q^2_1 r_{21} =0, \, r_{11} q^2_1 - q^1_1 r_{21} =0,\\
&2 \widehat r_{11} \widehat q^1_1 - \widehat r_{21} \widehat q^2_1 =0, \widehat r_{11} \widehat q^2_1 - \widehat r_{21} \widehat q^1_1=0.
\end{align}

To determine the holomorphic gauge kinetic coupling we need to consider the expansion of the holomorphic three-form
$\Omega$ before the orientifold projection. Therefore we write $z^3 = x^5 + \tau x^6$ where $\tau$ is the complex
structure modulus that will get fixed by the orientifold projection to $\tau=\ic$. If we keep the real three forms as
defined above in term of the $dx^i$ then we find (here we are using our freedom to not choose a normalization so that $\mathcal{Z}^0$ is as before)
\begin{align}
\Omega &= \frac{\sqrt{2}}{(1-\ic\tau)}dz^1 \w dz^2 \w dz^3 = \mathcal{Z}\mathcal{A}-\mathcal{F}\mathcal{B}+\mathcal{X}a-\mathcal{G}b\\
&=  \frac{1}{\sqrt{2}}\lp  \lp\mathcal{A} + \ic \mathcal{B}\rp + \frac{(\ic -\tau)}{(1- \ic \tau)} \lp a-\ic b\rp \rp.
\end{align}
There is thus a very simple relation $\mathcal{G}=i\mathcal{X}$ for all $\tau$, and the electric gauge kinetic coupling is given by
\be
f=-\frac{i}{2}\p_\mathcal{X} \mathcal{G}|_{\mathcal{X}=0} =  \frac{1}{2}.
\ee

The D-term in our example in the gauge where all charges are electric i.e., have lower indices, is
\be
D = \frac{e^\phi}{2 \mathcal{V}_6}  \lp \lp 4\mathcal{V}_6+v^3\lp 2(u^1)^2+(u^2)^2\rp\rp s_1
+v^3\lp u^1q^1_1+u^2q^2_1\rp-v^1r_{11}-v^2r_{21}-v^3r_{31}\rp.
\ee
So the contribution from the D-term to the potential  is
\bea
V_D &=& \frac{1}{2} (\Re f)^{-1} D^2 \\
&=& \frac{e^{2\phi}}{4 (\mathcal{V}_6)^2} \lp \lp 4\mathcal{V}_6+v^3\lp 2(u^1)^2+(u^2)^2\rp\rp s_1
+v^3\lp u^1q^1_1+u^2q^2_1\rp-v^1r_{11}-v^2r_{21}-v^3r_{31}\rp^2.\non
\eea

Finally we need to consider the tadpole constraints in this model.  We will start with a more general (but very brief) discussion of orientifold tadpoles.  Recall that the elements of a general orientifold group $G$ is a $\Z_2$ extension of an orbifold group $H$,
\be
1\longrightarrow H\longrightarrow G\longrightarrow \Z_2\longrightarrow 1,
\ee
where the elements of $G$ that are not in the image of $H$ are to be paired with the worldsheet parity operator $\Om_p$ (and possibly a factor of $(-1)^{F_L}$).  Put more simply, we can find a spacetime symmetry $\s$ such that
\be
G=H\cup \lp H\s\Om_p\rp,
\ee
and we require that $H$ be a group of spacetime symmetries, and that $\lp H\s\rp^2\subseteq H$.  There  is a twisted sector of states for each element $h\in H$, but no twisted sectors corresponding to the elements $h\s\Om_p$.  For the example at hand, $H=\langle\Theta\rangle=\Z_4$, and $\s$ is as given in equation (\ref{ExampleInvolution}), with $\s^2=\Theta$.  Each element $h\s\Om_p$, $h\in H$, generates a tadpole, via a crosscap diagram, for a R-R field in the $(h\s)^2$-twisted sector, localized at $h\s$-fixed points.

Consider now the potential tadpoles in our example.  There are tadpole contributions to the $\Theta$-twisted sector from $\s\Om_p$ and $\s\Theta^2\Om_p$, and contributions to the $\Theta^3$-twisted sector from $\s\Theta\Om_p$ and $\s\Theta^3\Om_p$.  There are no untwisted-sector tadpoles, so we only need to worry about possible twisted-sector (or fractional) O-planes.  In fact, the relevant crosscap diagrams are computed in~\cite{Rabadan:2000ma,Aldazabal:1998mr}, and for this particular model it is shown that the two contributions to each twisted sector cancel.  There are no localized tadpoles in this model to worry about, and in particular no need to add any D-branes.  In this case we can choose a symplectic basis $(\mathcal{A},\mathcal{B})$ which preserves the form of the NS-NS fluxes above and in which we have the simple relation $F_3=e\mathcal{B}$ and the tadpole constraints are automatically satisfied.

With this we can very simply write down the superpotential (dropping the redundant subscript $1$)
\be
W=-\frac{1}{\sqrt{2}}\ls e+p\tau+r_aG^a+\widehat{q}^\al T_\al\rs.
\ee

\subsection{An O5/O9 example}

There is a closely related example of the O5/O9 type which also exhibits D-terms.  The construction is the same as above except that we take our involution to be $\s'=\s\mathcal{I}_6$, with $\s$ as in (\ref{ExampleInvolution}).  In this case the full orientifold group is $\Z_{8-I}$.

Of course there really isn't any new physics; this O5/O9 construction is in fact precisely T-dual to the O3/O7 construction above, by dualizing the $x^5$ and $x^6$ coordinates.  For this reason the tadpoles also continue to cancel.

\section{Conclusions}
\label{Conclusions}

In this paper we have tried to illustrate how generalized NS-NS fluxes in type II orientifold compactifications can enrich the structure of the four-dimensional effective field theory.  In particular, we have shown how these fluxes can act as electric and magnetic charges for the R-R axion fields in four-dimensions, thus giving rise to D-term contributions to the scalar potential.  The hope is that these extra contributions to the potential will make it more likely to find interesting vacua.  It would be very interesting, for example, to repeat the exercise of~\cite{Hertzberg:2007ke} with these extra ingredients added.  It would also be nice to use the D-terms to find de Sitter minima of the potential.  Unfortunately, because of the relationship (\ref{GeneralDTerm}) between D-terms and F-terms, we can not use D-terms to uplift an otherwise supersymmetric vacuum, at least perturbatively (but see~\cite{Achucarro:2006zf} for a suggested nonperturbative effect).

Nearly all of the discussion in this work has been at the level of effective field theory, so it is very difficult to know which models can really be obtained from ten-dimensional string theory constructions, and in which regimes we can trust the approximations that we have been making, i.e. that the supergravity analysis (perhaps augmented by dualities) holds, that Ka{\l}u\.za-Klein modes are heavy enough to be ignored, and that backreaction of fluxes and localized sources, especially orientifold planes, can be kept under control.  These issues deserve a much more detailed exploration which we will not provide in the current work, though some relevant comments can be found in~\cite{DeWolfe:2005uu,Banks:2006hg,Ihl:2007ah}.

One approach to answering the question of which models can be obtained from well-defined ten-dimensional constructions, at least for toroidal orientifolds, is the base-fiber approach described for IIA in~\cite{Ihl:2007ah}, and following the spirit of~\cite{Dabholkar:2002sy}.  These techniques can easily be carried over to type IIB (or the heterotic string, for that matter), and for a given toroidal orientifold, one could identify which classes of fluxes could be constructed using these methods.  These constructions have the advantage of revealing the correct quantization conditions for the generalized NS-NS fluxes, which turn out to be non-trivial in general.  For configurations of fluxes which are not constructible in this way, it is not clear what quantization conditions are correct, or indeed even if the configurations themselves have a ten-dimensional origin.  Even when we do understand the NS-NS quantization, there is still some mystery about the R-R quantization conditions, which would require a better understanding of the relevant K-theory for these spaces~\cite{BergmanRobbins}.

This is a vexing situation since, as we have seen above, the effective theory structure actually fits together very nicely, and looks as though it could be applied to general Calabi-Yau orientifolds, rather than just toroidal examples.  Unfortunately, besides the confusions about quantization conditions, it also seems to be difficult to get the full set of Bianchi identities from geometric data in the general case.  Between the quantization conditions (which can sometimes have no nontrivial solutions) and the extra Bianchi identities, it seems likely that these general models will be much more constrained than they might naively appear.  However, it is our opinion that this shouldn't necessarily discourage attempts to use these effective theories to construct phenomenologically interesting scenarios.

\section*{Acknowledgements}
It is a pleasure to thank Aaron Bergman, Jacques Distler, Simeon Hellerman, and Matthias Ihl for many helpful comments, discussions and encouragement. The research of the authors is based upon work supported by the National Science Foundation under Grant No. PHY-0455649.

\providecommand{\href}[2]{#2}\begingroup\raggedright\endgroup


\bibliographystyle{utphys}
\end{document}